\pdfoutput=1

\documentclass[journal]{IEEEtran}
\usepackage{graphicx} 
\usepackage{caption} 

\usepackage{subfig} 
\usepackage{tikz}
\usepackage{cite}
\usetikzlibrary{shapes.geometric, arrows}
\usepackage{array}
\usepackage{textcomp}
\newcolumntype{C}[1]{>{\centering\arraybackslash}m{#1}}
\usepackage{relsize}
\usepackage{stackengine}
\usepackage{comment}
\usepackage[font={footnotesize}]{caption}
\usepackage{amsmath,amssymb,amsfonts,accents}
\newcommand{\myvect}[1]{\accentset{\rightharpoonup}{#1}}
\usepackage{algorithmic}
\usetikzlibrary{calc,positioning}
\usepackage{amsmath}
\usepackage{subfig} 
\usepackage{tikz}
\usepackage{cite}
\usetikzlibrary{shapes.geometric, arrows}
\usepackage{array}
\usepackage{textcomp}
\newcolumntype{C}[1]{>{\centering\arraybackslash}m{#1}}

\usepackage{threeparttable}

\usepackage{multirow}
\usepackage{hhline}

\newcolumntype{P}[1]{>{\centering\arraybackslash}p{#1}}
\newcolumntype{M}[1]{>{\centering\arraybackslash}m{#1}}
\usepackage[switch]{lineno}

\ifCLASSINFOpdf
\else
\fi
\begin{document}
%

\title{Multi-Material Blind Beam Hardening Correction Based on Non-Linearity Adjustment of Projections}




%
%
%

\author{Ammar Alsaffar, Kaicong Sun, Sven Simon}

\maketitle

\begin{abstract}

Beam hardening (BH) is one of the major artifacts that severely reduces the quality of Computed Tomography (CT) imaging. In a polychromatic X-ray beam, since low-energy photons are more preferentially absorbed, the attenuation of the beam is no longer a linear function of the absorber thickness. The existing BH correction methods either require a given material, which might be unfeasible in reality, or they require a long computation time. This work aims to propose a fast and accurate BH correction method that requires no prior knowledge of the materials and corrects first and higher-order BH artifacts. In the first step, a wide sweep of the material is performed based on an experimentally measured look-up table to obtain the closest estimate of the material. Then the non-linearity effect of the BH is corrected by adding the difference between the estimated monochromatic and the polychromatic simulated projections of the segmented image. The estimated monochromatic projection is simulated by selecting the energy from the polychromatic spectrum which produces the lowest mean square error (MSE) with the acquired projection from the scanner. The polychromatic projection is estimated by minimizing the difference between the acquired projection and the weighted sum of the simulated polychromatic projections using different spectra of different filtration. To evaluate the proposed BH correction method, we have conducted extensive experiments on the real-world CT data. Compared to the state-of-the-art empirical BH correction method, the experiments show that the proposed method can highly reduce the BH artifacts without prior knowledge of the materials.
\end{abstract}


%
\IEEEpeerreviewmaketitle

\section{Introduction}
\IEEEPARstart{C}{omputed} Tomography (CT) is an imaging technique that has been widely applied in medical and industrial applications such as medical diagnosis and non-destructive testing. However, CT usually suffers from undesirable artifacts such as beam hardening (BH), scattering, motion, cone beam, ring, and metal artifacts due to the physical nature of the X-rays \cite{Boas}. One of the major artifacts that influence the image quality is the BH artifact, which results from the polychromatic nature of the X-ray source. When photons of different energies penetrate through the object, low-energy photons are more easily absorbed than high-energy photons. This shifts the mean of the spectrum to a higher value and results in what is known as the hardening of the beam. Since the attenuation value of the material is energy-dependent, it decreases as the beam is hardened. This results in a non-linearity between the attenuation of the beam and the length of the propagation path. Applying a linear reconstruction method such as the filtered back projection (FBP) method, which assumes that the object attenuates the X-ray linearly with the path length, results in cupping and streak artifacts which degrade the quality of the reconstructed image.

Several methods were implemented to correct the BH artifact, such of these methods are filtration, dual-energy, linearization, post-reconstruction, statistical polychromatic reconstruction, and iterative reconstruction methods \cite{Gompel}, \cite{Krumm}. Although it is simple and widely applied, the filtration method is only able to reduce the BH artifacts and does not perform well in multi-material applications \cite{Brooks},\cite{McDavid}. The Dual-energy method requires two scans, this is in general expensive. In addition, it is limited to certain applications \cite{Gompel}. On the other hand, the linearization and the post-reconstruction methods require the material to be known which is not always possible. Although some linearization methods do not require the material knowledge \cite{Casteele,Gao,Mou}, they have some limitations. These methods are either limited to a certain class of objects or a specific geometry \cite{Gompel}. Moreover, the main burden of the statistical and the iterative reconstruction methods is the long computation time required,  which makes them not suitable for industrial applications \cite{Reiter}. In this work, a multi-material blind BH correction algorithm is proposed. This method does not require prior knowledge of the materials and can correct first order (cupping) and high order (streak) BH artifacts in multi-material objects by accurately mapping the polychromatic projection to an equivalent monochromatic one through a correction term that is added to the original projections from the scanner. All the simulations of the projections, which are needed for the calculation of the correction term in the proposed BH correction method, are performed using an in-house implemented and extensively verified multi-GPU accelerated Monte Carlo (MC) photon forward projection model \cite{Alsaffar}. This makes the overall process of the BH correction fast. The proposed method has been evaluated by several experimental examples from real-world data. The algorithm results in a fast and accurate BH correction which makes it very promising for medical and industrial applications.

\subsection{Contributions}

\subsubsection{Fast and Robust Material Estimation Based on Look-up Table}

To overcome the limitation of many other BH correction methods, which require the knowledge of the materials used, we have created a look-up table that contains experimentally measured polychromatic linear attenuation values of a wide range of well-known materials versus the polychromatic energy. Based on the look-up table and the measured linear attenuation value of each material inside the uncorrected volume, the proposed algorithm can iteratively find the best matched material by means of minimum mean square error (MMSE). Experimental results show that the proposed material estimation is robust to strong BH effects and a small deviation of the estimated material has a negligible impact on the BH corrected results.


\subsubsection{Accurate Regression of the Measured Projections by Aggregation of Weighted MC Simulations} 

To correct the non-linearity of the acquired projections, a correction term is required. One determinant part of the correction term is the estimated polychromatic projection. Estimation of such projection using MC simulators leads to an inaccurate result since the linear attenuation tables of the materials used in such simulators are not very precise, the same is also applied for the polychromatic spectrum used. To enable an accurate fitting to the measured data, the estimated projection is constructed by a linear combination of eight simulated polychromatic projections of different spectra by using diverse filtration. It is shown that the constructed projection very well matches the raw data.


\subsubsection{Estimation of Monochromatic Projection}

The other part of the correction term is the estimated monochromatic projection. This monochromatic projection is simulated by the opted energy bin which produces the lowest mean square error (MSE) with the acquired projection. While other methods used to determine this energy require information that is sometimes not available \cite{Millner}, the proposed method provides an accurate and fast estimation of the effective energy of the polychromatic spectrum using only the materials estimated from the proposed material estimation method.

\subsection{Related Work}

 In the literature, the methods of BH correction can be classified into 6 categories which are based on hardware filtration, dual-energy, linearization, post-reconstruction, statistical polychromatic reconstruction, and iterative reconstruction \cite{Gompel}, \cite{Krumm}. Some other methods combined two classes to produce an effective correction method.
 
 The first approach implies the use of filters \cite{Hsieh,Hunter} in which a thin plate of metal, aluminum, copper or other material is placed between the source and the object to filter out the low energy photons in the spectrum before their penetration through the object. This method has been widely used in the CT field due to its simplicity and the limited requirement of this method. However, it is not completely able to eliminate this artifact especially in the case of multi-material objects with different high attenuation characteristics \cite{Brooks},\cite{McDavid}. Besides, The use of these kinds of filters results in a lower detected signal to noise ratio and reduces the contrast between the materials
 \cite{Thomsen}.
 
 
 In the dual-energy methods \cite{Brooks}, \cite{Alvarez}, two images from the low and high energies are acquired. Then a material decomposition is performed by representing the linear attenuation coefficient by two bases functions, i.e., Compton scattering and photoelectric absorption \cite{Casteele,Alvarez,Macovski}. The contribution from these two bases function is calculated which enables the reconstruction of a monochromatic image. The main flaw in this method is the need to acquire two separate distinct measurements which is in general expensive.

The linearization methods \cite{Brooks}, \cite{Casteele}, \cite{Hammersberg, Herman} map the polychromatic projection to a monochromatic one through the approximation of the polychromatic projection as a curved function of the attenuation values versus the thicknesses traveled in the object. This function should be then mapped into a linear one which represents the monochromatic projection. A polynomial fitting is performed to derive the corrected projection. For small BH artifacts, the fit of a second-order polynomial is enough \cite{Casteele}, \cite{Herman}. However, for large BH artifacts, the fit of a higher-order polynomial is required \cite{Casteele}. For multi-material objects, the first reconstructed data is used as prior information about the scanned object which provides information about the intersection of each material with each ray path. The linearization can be then performed using the iterative post reconstruction (IPR) approach. The corrupted volume can be then corrected by incorporating the knowledge of the spectrum, the materials, and the material thickness. The first corrected reconstructed image is then used in the next iteration to derive a better-corrected image. A major drawback of this method is that the materials should be known. However, some of the linearization methods require no prior knowledge about the materials \cite{Casteele,Gao,Mou} but they have certain limitations, which includes the limitation to a certain class of objects or a specific geometry \cite{Gompel}. In \cite{Krumm} the authors proposed a BH correction method that is iterative in nature and similar to the IPR methods. The proposed method does not require the materials knowledge nor the spectrum and uses a hypersurface and a hyperplane to represent the polychromatic and the monochromatic attenuation values plotted with respect to the ray path through each material. The difference between the hypersurface and the hyperplane is calculated which is then added to the original projection from the scanner to correct the BH artifact. This method works well in case a proper segmentation can be performed on the original uncorrected volume \cite{Gompel}.

In the post-reconstruction methods, the difference of the monochromatic and the polychromatic projections estimated from a known spectrum, material density, mass attenuation, and atomic number is calculated and then added to the original projection from the scanner to correct the corrupted projections \cite{Nalcioglu}. The corrected projections are then reconstructed by the FBP method to produce a BH-free image \cite{Krumm}. The polychromatic projections estimation is done by first estimating the spectrum, this could be done by weighting spectra acquired using different filtration. The optimum spectrum should produce the lowest MSE between the projection calculated using this spectrum and the original projection from the scanner. Then the calculated projection from the estimated spectrum is subtracted from the derived monochromatic projection to form the correction term \cite{Zhao}. Very promising results have been acquired using this method.

Statistical correction methods \cite{Elbakri,Elbakri2,Cai,Wu,Yang,Gu} exploit multiple information to correct the BH artifacts in the X-ray images. Such information include the source and the detector model, the noise distribution, the measurement non-linearity, and the scatter effect are all used in the statistical methods in which they all incorporated into the maximum likelihood (ML) algorithm \cite{Gompel}. The major limitation in the statistical approaches is the long computational time required.

On the other hand, iterative algorithms improve the images iteratively by minimizing the cost function formed from the difference between the measured polychromatic images from the scanner and the simulated one. The spectrum, which is already involved in the minimization method along with the materials attenuation values, is approximated by a small number of energy bins \cite{Gompel}. In \cite{Yan} the authors proposed a new reconstruction algorithm which corrects the BH artifacts by involving the polychromatic characteristic of the X-ray beam. The algorithm in \cite{Man} corrects the BH effect by considering a polychromatic acquisition model and taking into account the energy dependency of the materials. In this method, the linear attenuation value of the material is decomposed into a photoelectric component and a Compton scatter component which are weighted based on a prior assumption. Moreover, the authors in \cite{Menvielle} proposed a method to reduce the BH artifacts by again incorporating the polychromatic nature of the X-ray source into account. A likelihood-based estimator is then used to determine the attenuation of the materials. The simultaneous algebraic reconstruction technique (SART) is the base of the work in \cite{Brabant}. The authors have used this technique to reconstruct a BH artifact-free image by incorporating this kind of artifact in the forward projection model of this reconstruction method. A model-based BH correction algorithm has been introduced in \cite{Jin}. This algorithm neither requires the X-ray spectrum nor the attenuation values of the materials. A polynomial function of both low and high-density materials is first formulated which represents a poly-energetic X-ray forward projection model. Then an alternating minimization algorithm is then developed which is used to estimate the reconstructed image, the material segmentation, and the two coefficients of the polynomial function which models the BH function.


Recently, the energy-sensitive detector also known as a photon-counting detector is introduced. These kinds of detectors provide multiple information about each detected photon including the energy of this photon. In \cite{Schumacher} it is shown that simple energy thresholding applied in such a detector allows the reduction of the BH effect and enhances the quality of the CT images.

\begin{figure}[ht]
\centerline{\includegraphics[width=0.8\linewidth]{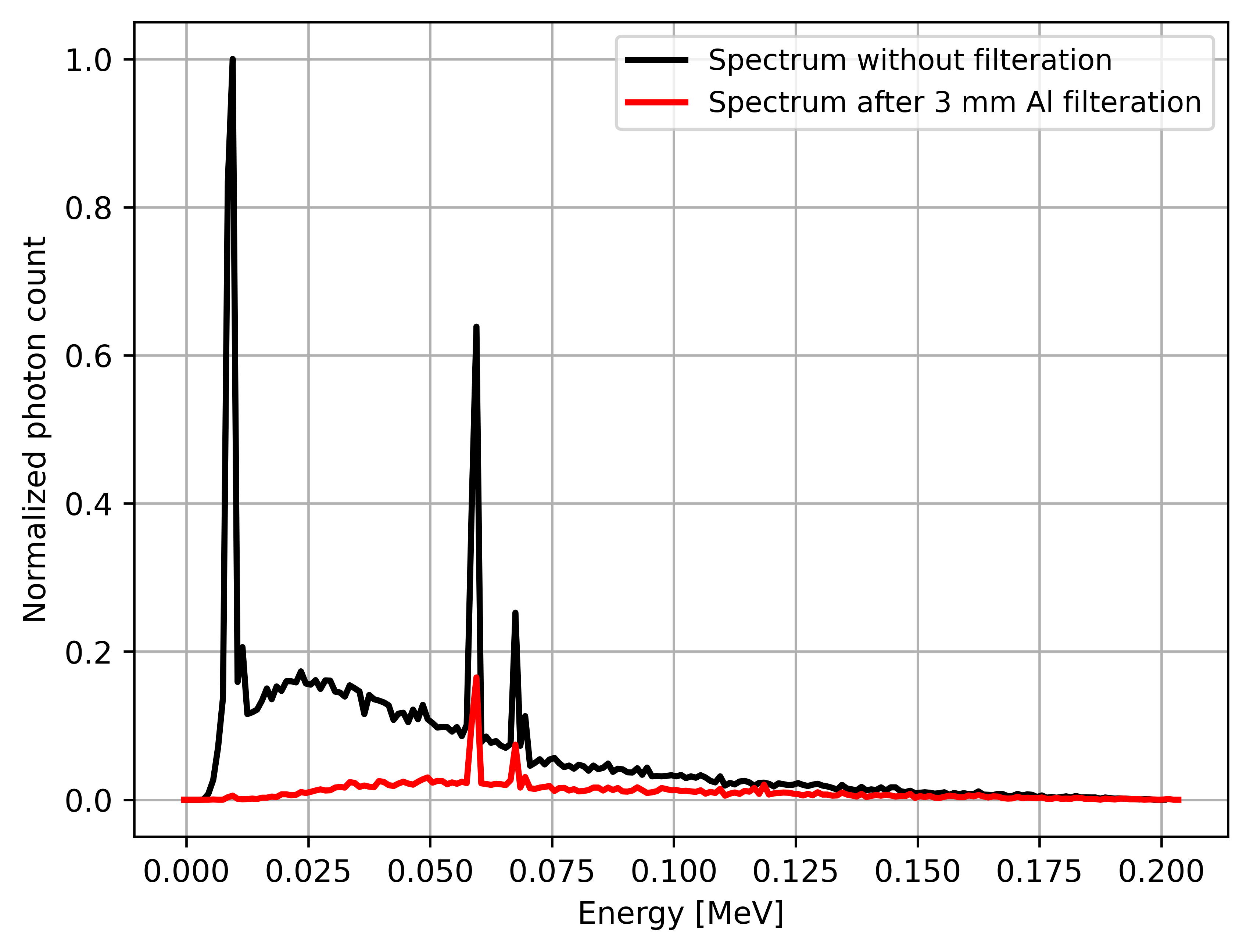}}
\caption{Hardening of the X-ray spectrum after passing through $3\,mm$ aluminum filter simulated using the Geant4 MC simulator \cite{GEANT4}, the black curve represents the spectrum of the X-ray source before the filter, while the red curve shows the spectrum of the source after passing through the filter.}
\label{spectrum}
\end{figure}

\begin{figure*}[ht]
\centerline{\includegraphics[width=\textwidth]{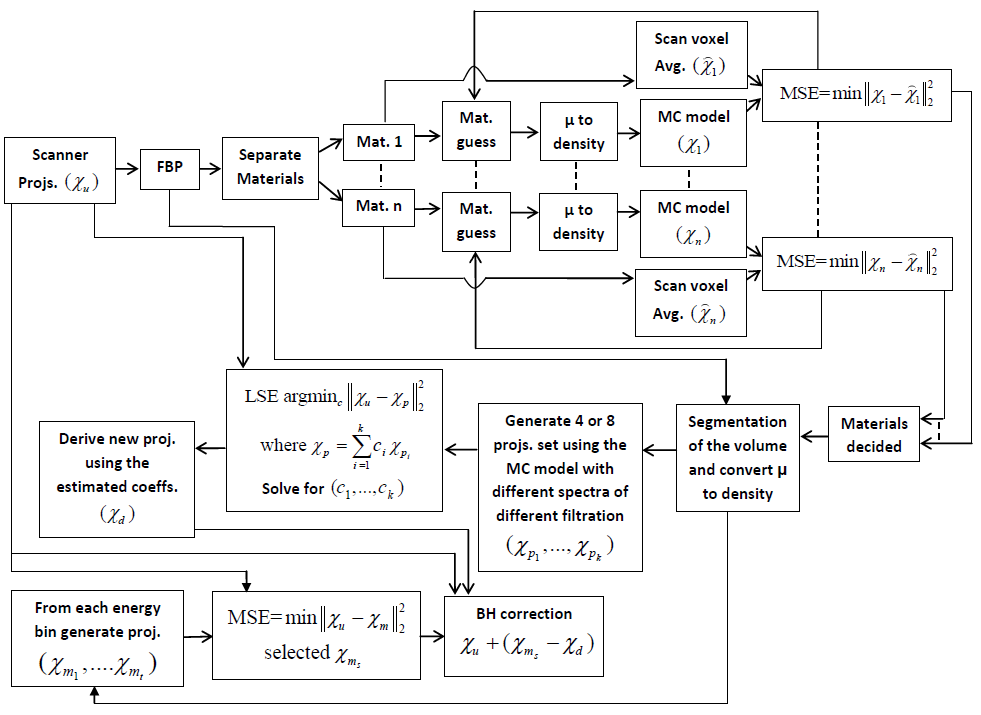}}
\caption{The flowchart of the proposed BH correction algorithm.}
\label{flowchart}
\end{figure*}

\section{Material AND Methods}

\subsection{Description of the Beam Hardening Problem }

Multiple material specifications influence the attenuation of the source intensity $I_0$. Such specifications include the material atomic number, the density of the material, the energy of the photon, and the source-detector propagation path. According to the Beer-Lambert law, the intensity on the detector from a monochromatic source is given by (\ref{eqmonochromatic}) \cite{Zhao},

\begin{equation}
I_{mono}=N\eta(E) exp\Big(-\int^L_0\mu(E,\myvect{x},\rho)d{\myvect{x}}\Big),\label{eqmonochromatic}
\end{equation}
$N$ is the number of incident photons, $\eta(E)$ is the energy response of the detector at energy $E$, $\rho$ is the density of the material, $\mu(E,\rho)$ is the energy-dependent linear attenuation coefficient, $L$ is the propagation path length, and $\myvect{x}$ is spatial position. For monochromatic source, the intensity of the rays has the same amount of attenuation for every slice of thickness $dx$ as in (\ref{eqmonochromatic2}),

\begin{equation}
dI/I=-\mu dx, 
\label{eqmonochromatic2}
\end{equation}

 This is because there is no change in the energy in such a source and the attenuation value $\mu(E,\rho)$, which is energy-dependent, does not change.

However, the source used by the real-world scanners is a polychromatic source in which the energy of the photons emitted by such a source varies over several energies. When the rays from such a source pass through an object, the ratio of the high to the low energy of the photons increases. This is because the low-energy photons easily got absorbed during their penetration and the mean of the beam is then shifted. Fig.~\ref{spectrum} shows how the mean of the spectrum is shifted due to the absorption of low-energy photons after passing through a $3\,mm$ aluminum filter. The attenuation of the materials inside the object, as its energy-dependent, will have then a lower value due to the increase in the mean energy. Thus the expression of the intensity from (\ref{eqmonochromatic}) is no longer valid. The intensity on the detector from such sources could be expressed by (\ref{eqpolychromatic}) \cite{Zhao},

\begin{equation}
I_{poly}=N\int^{Em}_0\eta(E)S(E)dE\,exp\Big(-\int^L_0\mu(E,\myvect{x},\rho)d\myvect{x}\Big).\label{eqpolychromatic}
\end{equation}
Here $Em$ and $S(E)$ are the maximum energy and the spectrum of the source respectively.

With the absence of the object, the monochromatic and the polychromatic intensities are given by \eqref{eqmonochromaticNoobject} and  \eqref{eqpolychromaticNoobject} respectively,
\begin{equation}
I_{mono,0}=N\eta(E).\label{eqmonochromaticNoobject}
\end{equation}
\begin{equation}
I_{poly,0}=N\int^{Em}_0\eta(E)S(E)dE.\label{eqpolychromaticNoobject}
\end{equation}
The non-linear relation between the intensity and the distance described in \eqref{eqpolychromatic}, causes the BH artifacts. Such non-linear relation could be compensated by finding a proper correction term which is able to map this non-linear relation to a linear one. A certain class of correction methods derives a correction term by calculating the difference between estimated monochromatic and polychromatic projections. This term is then added to the original projections to correct the BH artifacts.

If \eqref{logmono} and \eqref{logpoly} represent the logarithm of the monochromatic and the polychromatic projections respectively. 

\begin{equation}
\chi_{mono}=ln\Big(\frac{I_{mono,0}}{I_{mono}}\Big),
\label{logmono}
\end{equation}

\begin{equation}
\chi_{poly}=ln\Big(\frac{I_{poly,0}}{I_{poly}}\Big),
\label{logpoly}
\end{equation}
where $\chi_{mono}$ and $\chi_{poly}$ are the monochromatic and the polychromatic attenuation projections respectively. The correction term is then calculated as in \eqref{eqdifference}.

\begin{equation}
\Phi_{CT}=\chi_{mono}-\chi_{poly},
\label{eqdifference}
\end{equation}
where $\Phi_{CT}$ is the correction term. The correction of the projections from the scanner is then performed according to \eqref{correction}.

\begin{equation}
\chi_c=\chi_u+\Phi_{CT},
\label{correction}
\end{equation}
where $\chi_c$ is the corrected attenuation projection, $\chi_u$ is the original attenuation projection from the scanner.

\subsection{The Proposed BH Correction Method}
The flowchart of the proposed BH correction algorithm is shown in Fig.~\ref{flowchart}. This algorithm is divided into three parts. The first part is to estimate the materials of the uncorrected volume from the real-world scanner. The estimated materials from the first part are then used in the other two parts of the proposed algorithm to calculate the BH correction term given in (\ref{eqdifference}).

\subsubsection{Materials Estimation}
Most of the available BH correction methods require prior knowledge of the materials. Access to such information in many industrial cases is not always possible \cite{Gompel}. Most of the proposed methods that do not require this information, either follow an iterative correction approach which is timely expensive \cite{Reiter} or perform a linearization with some restrictions \cite{Gompel}. To overcome this limitation, the proposed BH correction method does not require prior knowledge of the materials. However, a simple and fast estimation of the materials is done to simulate the two projections in the BH correction term using the GPU-based MC model. Especially, a look-up table containing polychromatic linear attenuation values of a wide range of well-known materials such as copper, aluminum is measured experimentally for different energies.


Firstly, the acquired intensity projections are converted into linear attenuation projections. A volume is then derived from the linear attenuation projections using the FBP reconstruction method. The different materials in the reconstructed volume are separated and then assigned into different volumes of the same size as the original volume, i.e., each volume contains a single material. The segmentation is performed using thresholds obtained from the Otsu method \cite{Otsu}. For each volume, a similarity match between the measured attenuation value in this volume and the synthesized attenuation of all the materials in the look-up table is performed by MSE as formulated in \eqref{MMSE}.


\begin{equation}
mse=\mathop{\mathbb{E}}\Big[\parallel\mu-\widehat{\mu}\parallel_2^2\Big],\label{MMSE}
\end{equation}
where $\mu$ represents the linear attenuation value of the material in the separated volume, and $\widehat{\mu}$ represents the linear attenuation values of all the materials from the table for the used energy. The linear attenuation value of the material from the table that produces the MMSE is then selected as the initial guess. Based on this initial guess, a projection $(\chi)$ using the primary photons only without the scattered ones is simulated using the GPU-based MC forward projection model. The GPU-based MC forward projection model performs the simulation using a voxelized volume. Each voxel should have a density value of the guessed material. The simulation of this projection is done according to \eqref{eqpolychromatic}, \eqref{eqpolychromaticNoobject}, and \eqref{logpoly}. 

To evaluate the accuracy of the guessed material. Another projection $(\widehat{\chi})$ is derived using the volume which has the separated material. This projection is acquired by simulating rays from the source to each pixel on the detector. The linear attenuation values of all the voxels crossed by each ray are multiplied by the distance traveled through these voxels and then summed up to generate the pixel value associated with this ray. This is done according to \eqref{polyvoxel}. The simulated projection $(\widehat{\chi})$ using the second method is closer to reality since this projection is acquired using the original linear attenuation values of the volume and no assumption of the material is there. The projections simulated using the GPU-based MC model and the second method, i.e., $\chi$ and $\widehat{\chi}$ respectively, are then compared in terms of MSE.


\begin{equation}
\widehat\chi_{j_{l}}=\sum_{v=1} \mu_{v} d_v,
\label{polyvoxel}
\end{equation}
where $\widehat\chi_{j_{l}}$ is the value scored on the detector pixel $l$ derived from the volume which contains the separated material $j$, $\mu_{v}$ is the linear attenuation value of the voxel $v$, $d_v$ is the distance that the ray travel in voxel $v$. For a better material estimation, the projection which is simulated using the GPU-based MC model is simulated again for the rest of the available materials in the table which have higher linear attenuation values than the initially guessed material. The projections generated by the GPU-based MC model are compared with the projection generated from the second method in a MMSE sense. The material from the table that results in a MMSE is then selected as a final decision. The same process is then repeated to estimate the rest of the materials. 

The proposed method of material estimation produces an estimation of the material that best fits the linear attenuation value of the material in the original uncorrected volume from the scanner. It should be noted that it is not necessarily true that the material estimation will result in correct detection of the material which is originally used. This is especially true if the linear attenuation value inserted into the algorithm, which belongs to the uncorrected volume from the scanner, is far from the value assumed on the table for the same material due to for example to a stronger BH and scatter artifacts. This is because the scanned object using the scanner could have a wider diameter or different geometry setup than the object which is used to fill the polychromatic look-up table. This inaccurate material estimation could be compensated by the proposed method of the polychromatic projection estimation described in Section \ref{EstimatedPolyMethod}. Section \ref{robutness of the material estimation} discusses the robustness of the material estimation method and the effect of the wrong material estimation on the result of the BH correction.

\subsubsection{Simulation of the Polychromatic Projections}
\label{EstimatedPolyMethod}
As the materials in the original uncorrected volume from the scanner are determined using the proposed materials estimation method, the linear attenuation values of each material in this volume are replaced by the density value of this material. As mentioned before, this is done because the GPU-based MC model requires a volume with density values to simulate the projections used in the BH correction process. However, to derive a correction term, as given in \eqref{eqdifference}, that can correct the non-linearity of the acquired projections from the scanner, an accurate estimation of the simulated polychromatic projection is required. Even when the materials are accurately estimated, the simple estimation of such projection using MC simulators provides an inaccurate result that does not perfectly match the one from the scanner. This is mainly because the linear attenuation tables of the materials and the polychromatic spectrum used in such simulators are not very precise. In addition, no matter how accurate the simulators are, they are not able to simulate all the real effects in the actual scanners.
Thus, there will be always a small mismatch between the simulated polychromatic projection and the one acquired from the scanner. 


To construct an accurate match to the raw projection, we propose to perform a regression from eight simulated polychromatic projections by the GPU-based MC forward projection model. These eight projections are simulated using primary photons only and with different filtration under spectra of the same energy. The spectra are all simulated using SpekCalc software \cite{SpekCalc}. Then a linear least square estimation (LSE) is used to minimize the difference between the original uncorrected projection from the scanner and the weighted sum of the eight simulated projections. This is done according to \eqref{LSE} and \eqref{LSE_2}. The optimized coefficients are used to weight the eight simulated projections and construct the best estimation. 


\begin{equation}
c=arg\; min_{c} \mathop\parallel \chi_u-\chi_{p}\parallel_2^2,
\label{LSE}
\end{equation}
where $c$ is the coefficients to be derived, $\chi_u$ is the original attenuation projection from the scanner, and $\chi_{p}$ is the weighted sum given by \eqref{LSE_2},

\begin{equation}
\chi_{p}=\sum_{i=1}^{k} c_{i}\chi_{p_{i}},
\label{LSE_2}
\end{equation}
where $k$ is the number of the simulated polychromatic projections used in the calculation of the weighted sum, $\chi_{p_{i}}$ is the polychromatic projection $i$ simulated using the GPU-based MC model. Section \ref{robofPoly} discusses the robustness of this method and shows how it can estimate a polychromatic projection that is very close to the original one. Additionally, the comparison results between the estimated polychromatic projection and the original projections from the scanner are also shown for two examples.



\subsubsection{Simulation of the Monochromatic Projections}
\begin{figure}[ht]
\centerline{\includegraphics[width=0.8\linewidth]{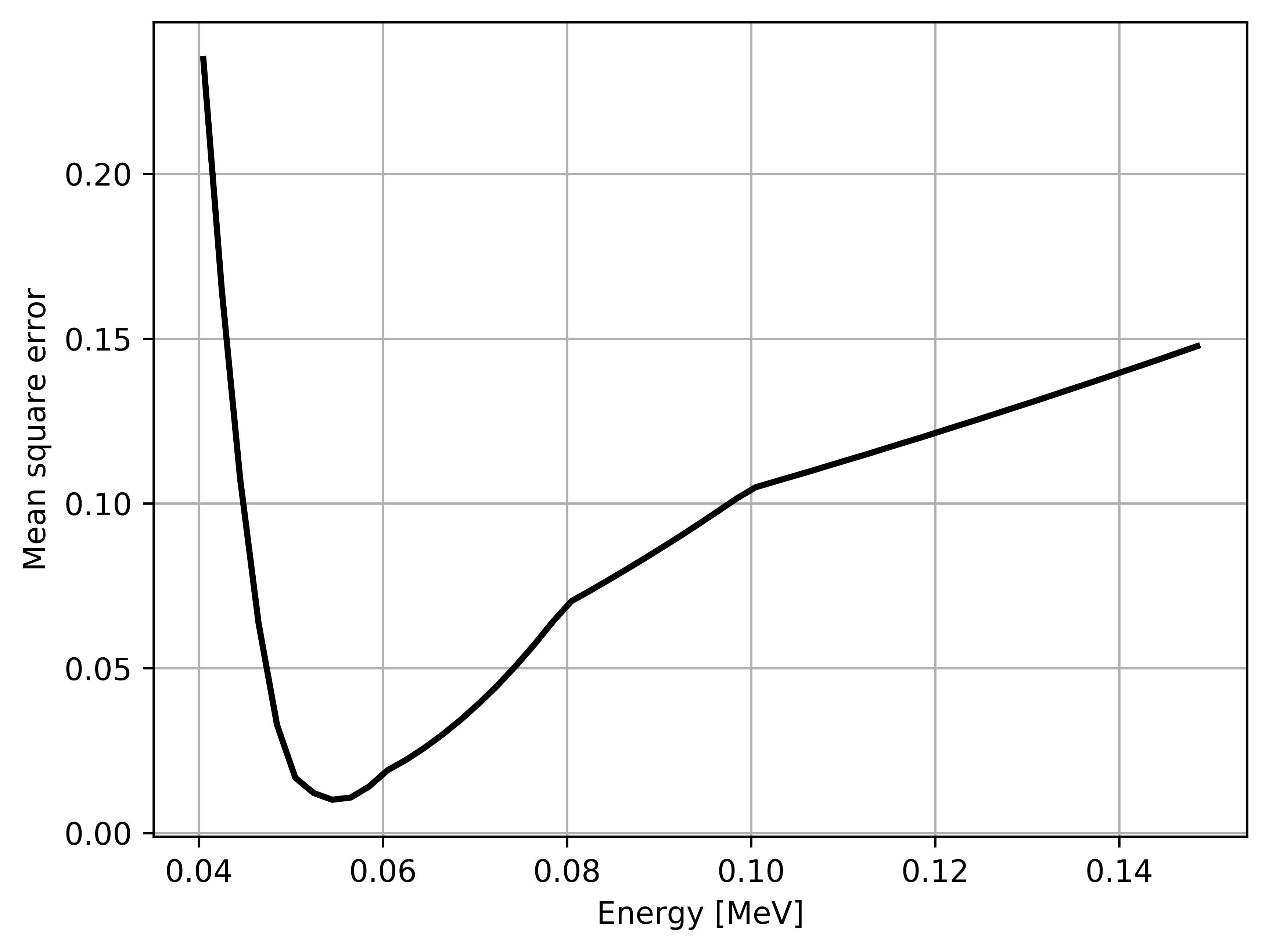}}
\caption{Plot of the MSE values calculated between the original projection from the scanner and the monochromatic projections from all the energy bins of the X-ray spectrum.}
\label{MMSEfigure}
\end{figure}

To complete the correction term, an estimation of the monochromatic projection is required. This monochromatic projection should be equivalent to the original polychromatic projection from the scanner. To simulate this projection, proper monochromatic energy has to be selected. Methods previously proposed to determine this energy require information that is sometimes not available \cite{Millner}. In this work, a fast and accurate method to determine this energy is proposed. The proposed method derives this energy by selecting each energy from the simulated polychromatic spectrum individually. The selected energy is used by the GPU-based MC forward projection model to simulate a monochromatic projection using \eqref{eqmonochromatic}, \eqref{eqmonochromaticNoobject}, and \eqref{logmono}. The simulated monochromatic projection is compared with the original projection from the scanner in terms of MSE. The energy that results in the lowest MSE is selected. Fig.~(\ref{MMSEfigure}) shows the results of the MSE between the simulated monochromatic projection from each energy bin and the original projection from the scanner. In this figure, it is shown that the energy which produces the lowest MSE is around the $55\,keV$, which is the energy selected to estimate the monochromatic projection. Section \ref{robofMono} discusses the robustness of the proposed method for estimating the monochromatic projection and shows how this method can estimate a monochromatic projection that is equivalent to the original polychromatic projection.

\subsubsection{Correction of the Projections}
\label{segflaws}
As the monochromatic and the polychromatic projections are estimated, the correction term can be calculated using \eqref{eqdifference}. This correction term is then added to the original acquired projection using \eqref{correction}. It is important to note that when the segmentation of the original volume from the scanner is optimum, the proposed estimation method of the polychromatic projection will be able to predict an accurate polychromatic projection that is very close to the one from the scanner. Consequently, the BH corrected projection will be close to the estimated monochromatic projection. However, in case that the segmented volume, from which the estimated polychromatic and the monochromatic projections are derived, is not perfect due to for example the severe BH and scatter artifacts or due to the inaccuracy of the segmentation method, the estimated polychromatic and monochromatic projections will be affected by this segmentation error. Nevertheless, this has a very limited impact on the performance of the proposed BH correction. We have shown more detailed insights in Section \ref{roboagainstseg}.   








\section{Experiments AND Results}

In this section, the comparison between the BH correction results of the proposed method and the state-of-the-art empirical BH correction method \cite{Kyriakou} is first presented. Then, extensive analyses of the proposed method were done. These include the effect of the wrong material estimation on the quality of the BH correction results and the robustness of the proposed methods of the polychromatic and the monochromatic projections estimations. Moreover, it is also shown how the proposed BH correction method can overcome possible segmentation errors. Finally, the scatter and its effect on the quality of the BH correction results are discussed.

\begin{figure*}[ht!]
\centerline{\includegraphics[width=0.9\linewidth]{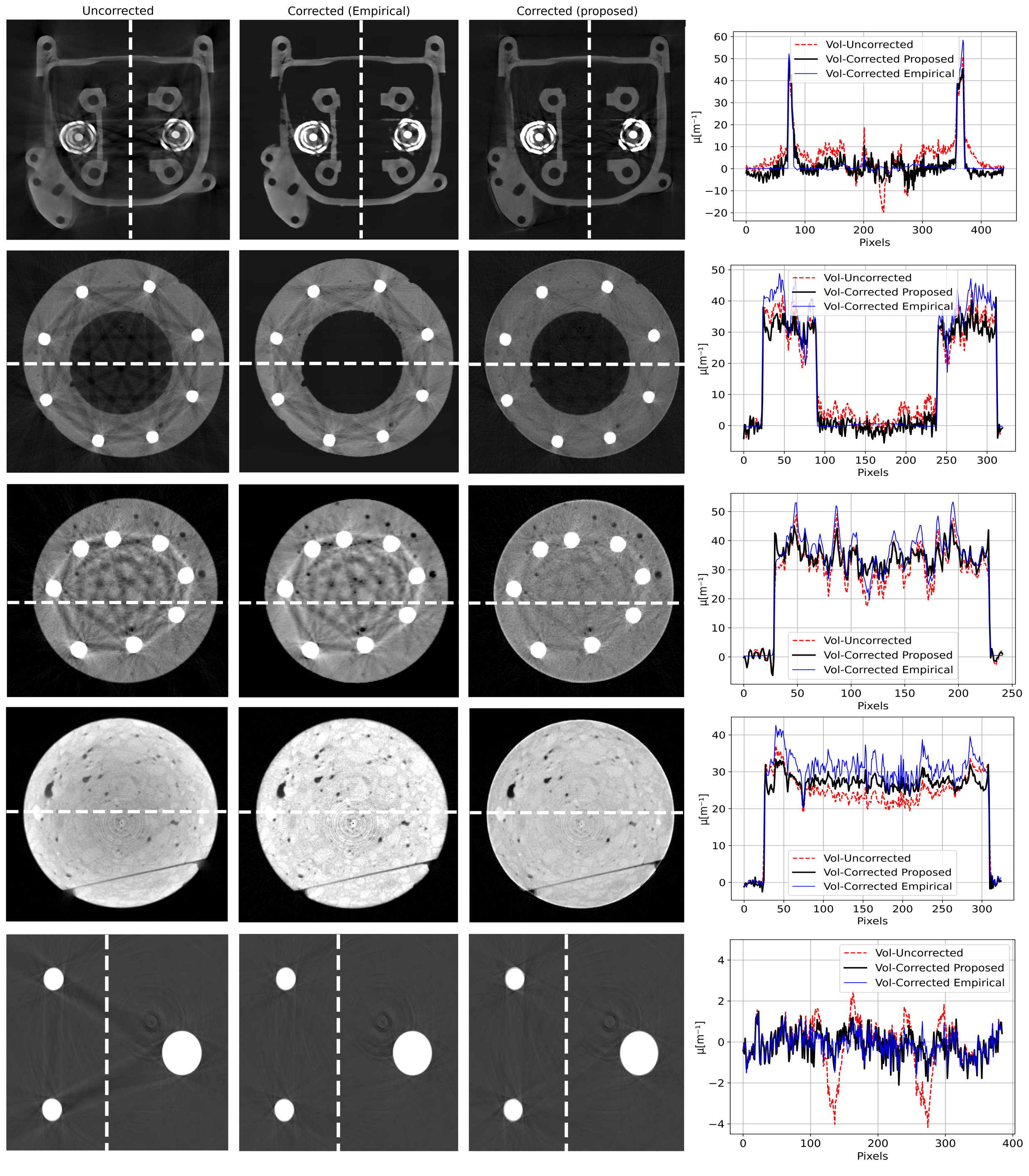}}
\caption{Experimental results of the proposed BH correction method. The first column shows the slices from the uncorrected volumes of five different objects. The second column shows the slices from the BH corrected volumes using the empirical method. The third column shows the slices from the BH corrected volumes using the proposed method. The last column shows the comparisons between the profile lines of the slices in columns 1 to 3. The object in the first row is made from aluminum and steel. The objects in the second and the third rows are made from cement and steel. The fourth object is a concrete cylinder. The last object represents three aluminum cylinders positioned away from each other in the space. The volumes which were used in the BH correction method are all near scatter-free volumes acquired by the use of the collimator. The simulated projections used in the proposed BH correction are from primary photons only without scatter. The profiles are marked by white dashed lines in these sub-figures.}
\label{Allexampleswithoutscatter}
\end{figure*}

\subsection{Evaluation of the Proposed BH Correction Method}

Fig.~\ref{Allexampleswithoutscatter} shows the BH correction results using the proposed method and the empirical BH correction method. From this figure, it is shown that the proposed BH correction method outperforms the empirical BH correction method. In the five presented examples, the proposed method can suppress both the cupping and the streak artifacts and show better visualization in the BH corrected volumes. On the contrary, the empirical method did not perform well in suppressing the streak artifacts within the same examples. The volumes from the scanner, shown in the first column of Fig.~\ref{Allexampleswithoutscatter}, are all near scatter-free volumes acquired by the use of a collimator.

\subsection{Extensive Analysis of the Proposed BH Correction Method}

\subsubsection{Effect of Inaccurate Estimation of Materials on BH Correction}

\label{robutness of the material estimation}
The proposed materials estimation method depends on the polychromatic linear attenuation values of the different materials in the uncorrected volume and the polychromatic linear attenuation values of the relevant materials in the look-up table. The linear attenuation values in the table were derived from fixed objects diameter, each object represents a different material. If the volume needs to be corrected is derived from the same object and geometry setup which were used to fill the table, the estimation of the material will be perfect. Otherwise, the linear attenuation values of the different materials in the uncorrected volume will deviate far from the ones assumed on the table. This deviation could be a consequence of a stronger BH and scatter artifacts. Thus, the proposed materials estimation method will wrongly estimate these materials. Due to the proposed estimation scheme of the polychromatic projection, even when the materials are wrongly guessed, this approach can estimate the polychromatic projection in a way that it will be close to the original polychromatic projection from the scanner. Thus the effect of the wrong material estimation could be compensated due to the robustness of the polychromatic projection estimation method. 

\begin{figure}[ht!]
\centering
\def\stackalignment{l}
\subfloat{\topinset{\bfseries \textcolor{white}{(a)}}{\label{fig:detectorefficiency}\includegraphics[width=0.4\linewidth]{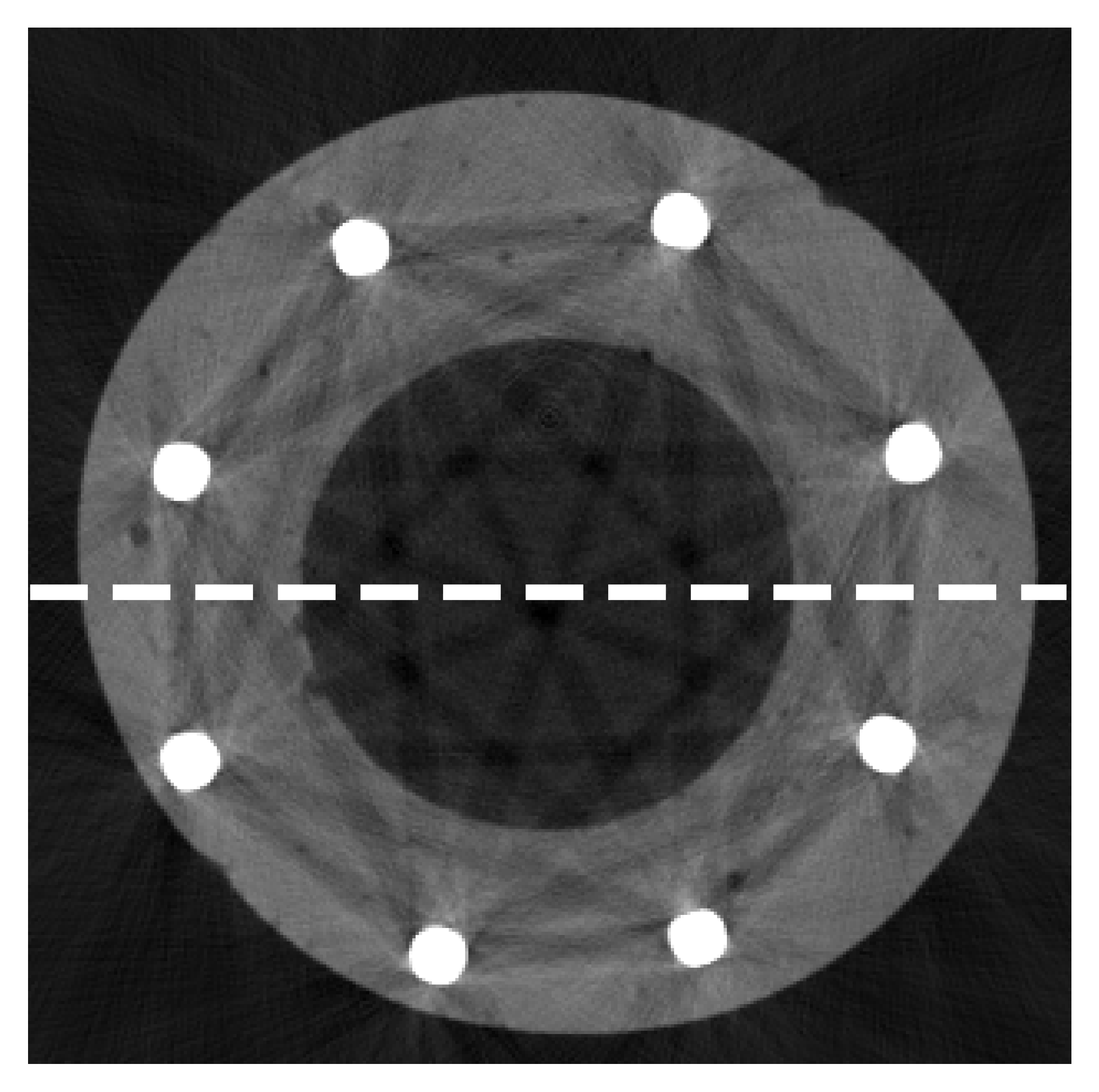}}{0.12in}{.1in}}\hspace{-0.4\baselineskip}
\subfloat{\topinset{\bfseries \textcolor{white}{(b)}}{\label{fig:CorrectlyEstimated}\includegraphics[width=0.4\linewidth]{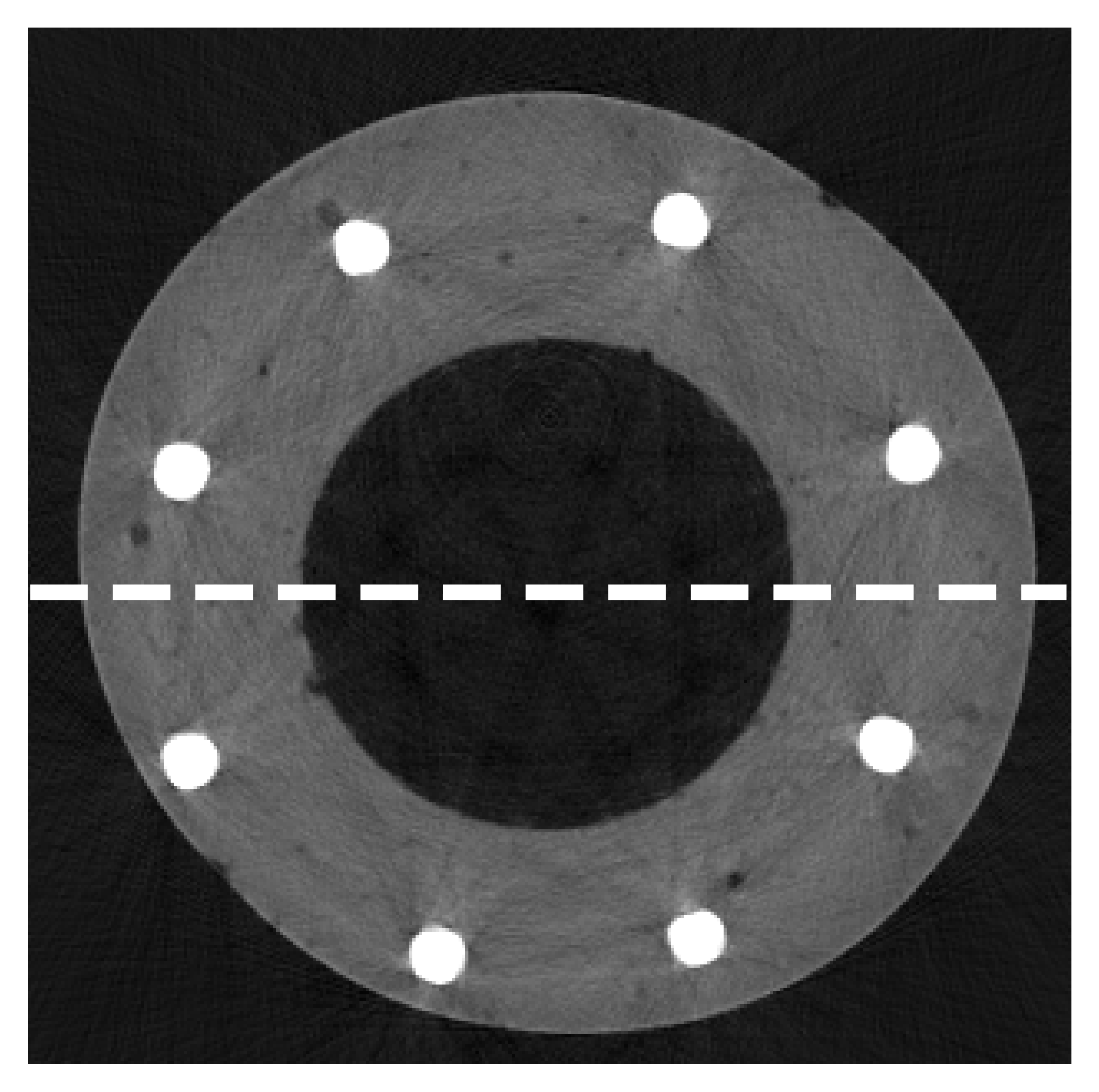}}{0.12in}{.1in}}\vspace{-1\baselineskip}
\subfloat{\topinset{\bfseries \textcolor{white}{(c)}}{\label{fig:LowerEstimated}\includegraphics[width=0.4\linewidth]{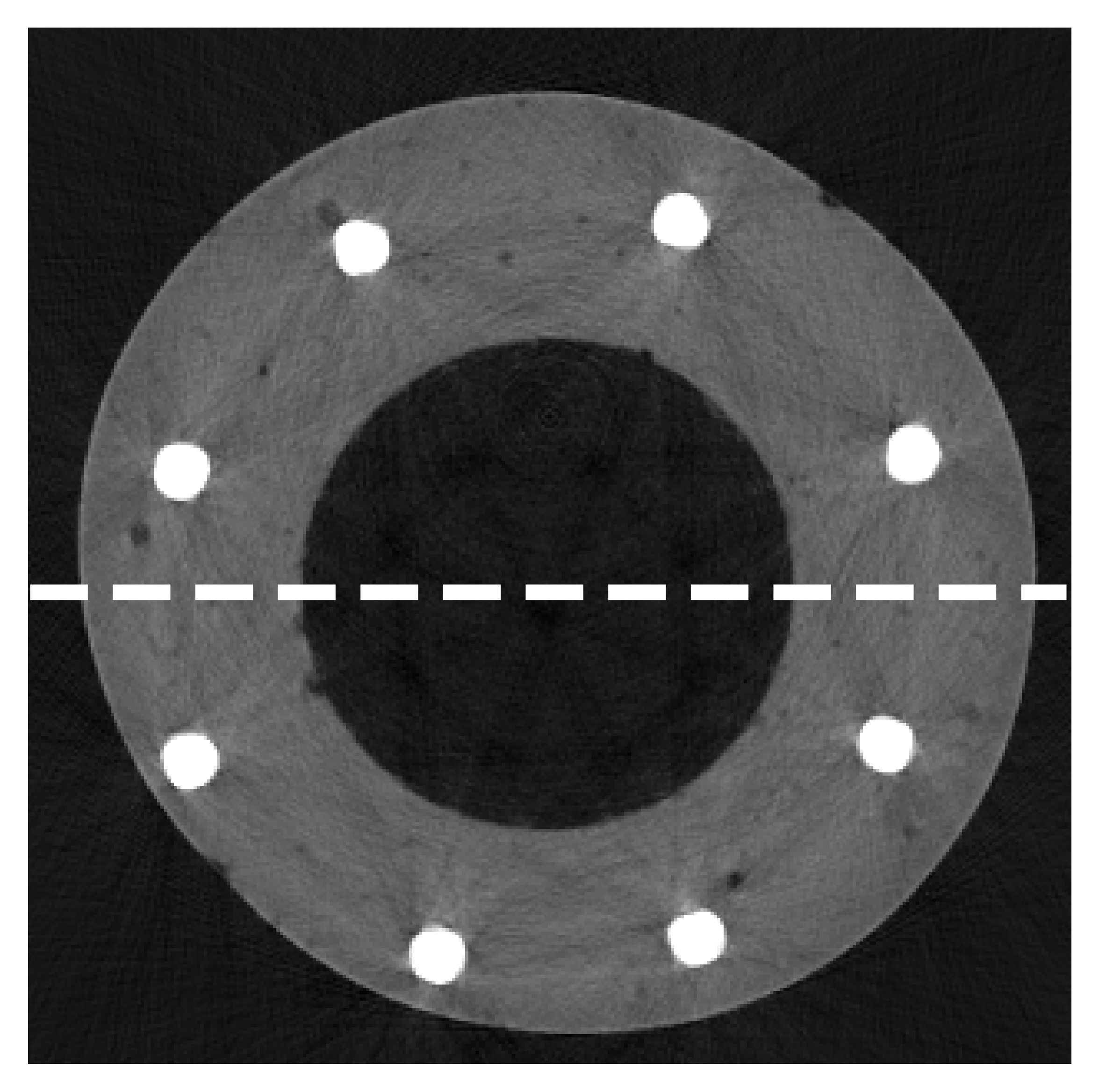}}{0.12in}{.1in}\hspace{-0.1\baselineskip}}
\subfloat{\topinset{\bfseries \textcolor{white}{(d)}}{\label{fig:UpperEstimated}\includegraphics[width=0.4\linewidth]{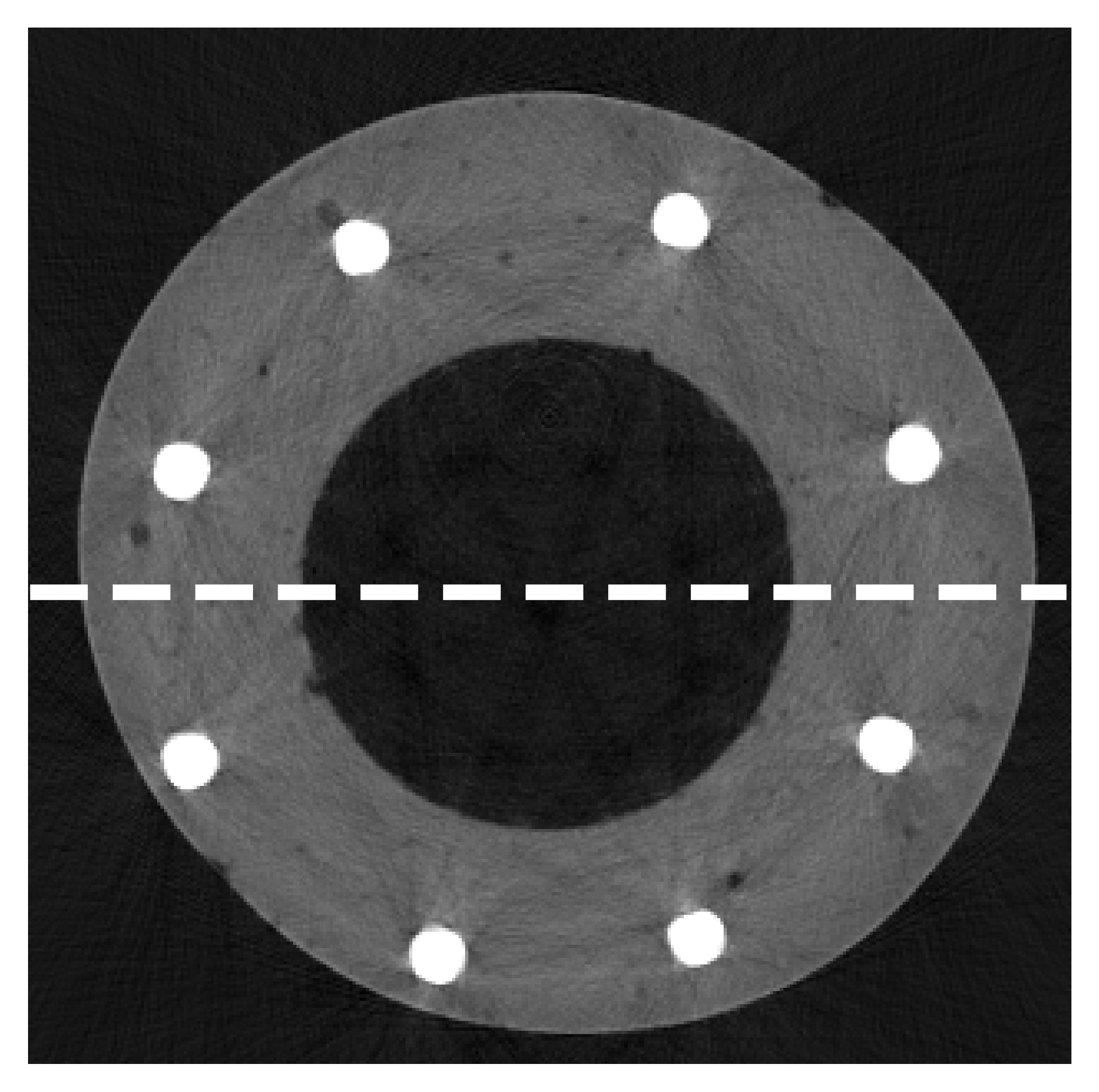}}{0.12in}{.1in}}\vspace{-0.1\baselineskip}
\subfloat{\topinset{\bfseries \textcolor{black}{(e)}}{\label{fig:depositenergy}\includegraphics[width=0.6\linewidth]{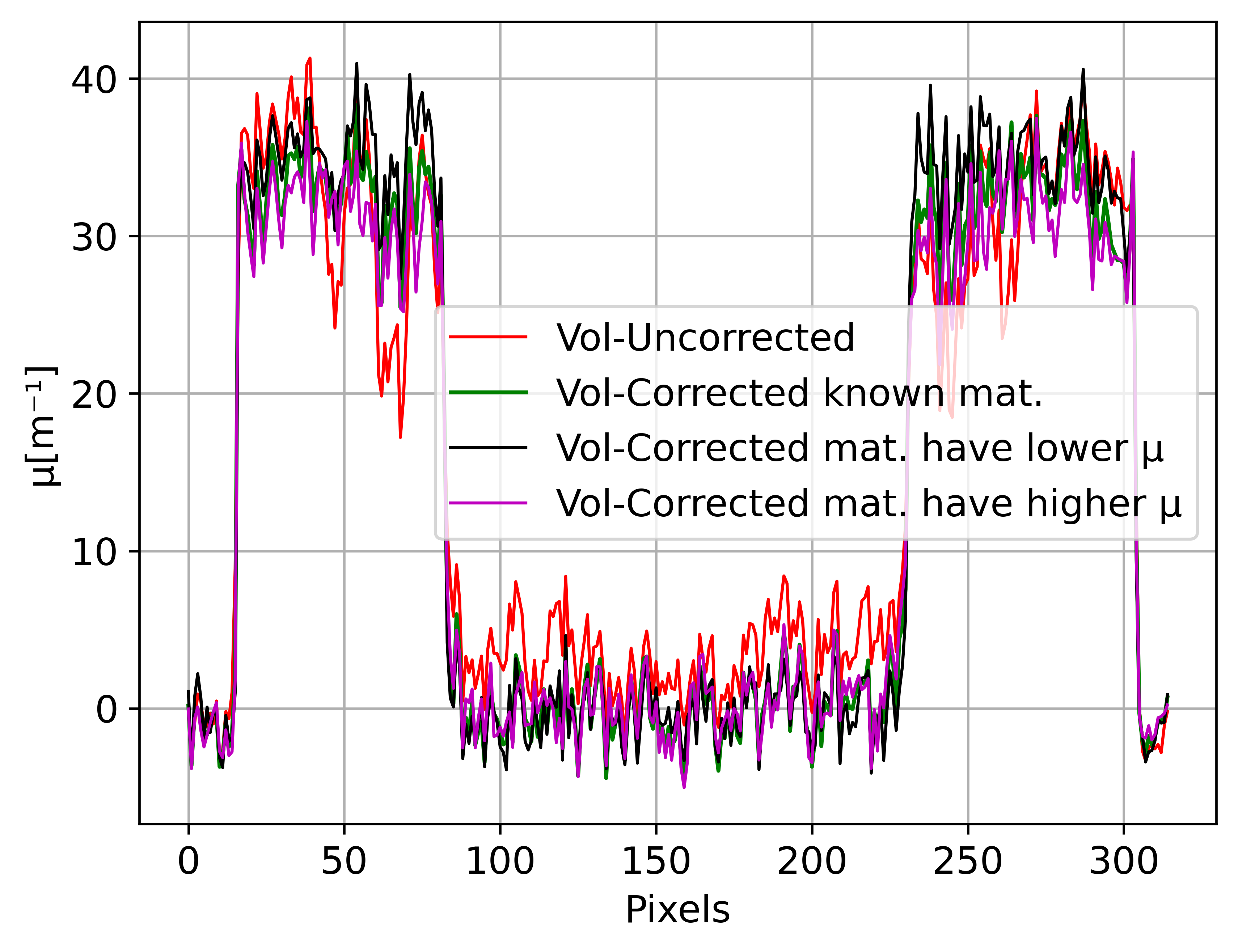}}{0.12in}{.27in}}\\

\caption{Results of the BH correction using the proposed method for different cases of materials estimation. (a) Slice from the uncorrected volume, (b) slice from the BH corrected volume assuming the materials are known, (c) slice from the BH corrected volume assuming the materials are wrongly estimated as silicon and titanium instead of the actual materials of cement and steel respectively, (d) slice from the BH corrected volume assuming the materials are wrongly estimated as aluminum and copper instead of the actual materials of the cement and steel respectively, (e) profile lines of the images in (a), (b), (c), and (d). The profiles are marked by white dashed lines in (a), (b), (c), and (d).}
\label{RobutOfMatEstim}
\end{figure}

\begin{figure}[ht!]
\centering
\def\stackalignment{l}
\subfloat{\topinset{\bfseries \textcolor{white}{(a)}}{\label{fig:detectorefficiency}\includegraphics[width=0.42\linewidth]{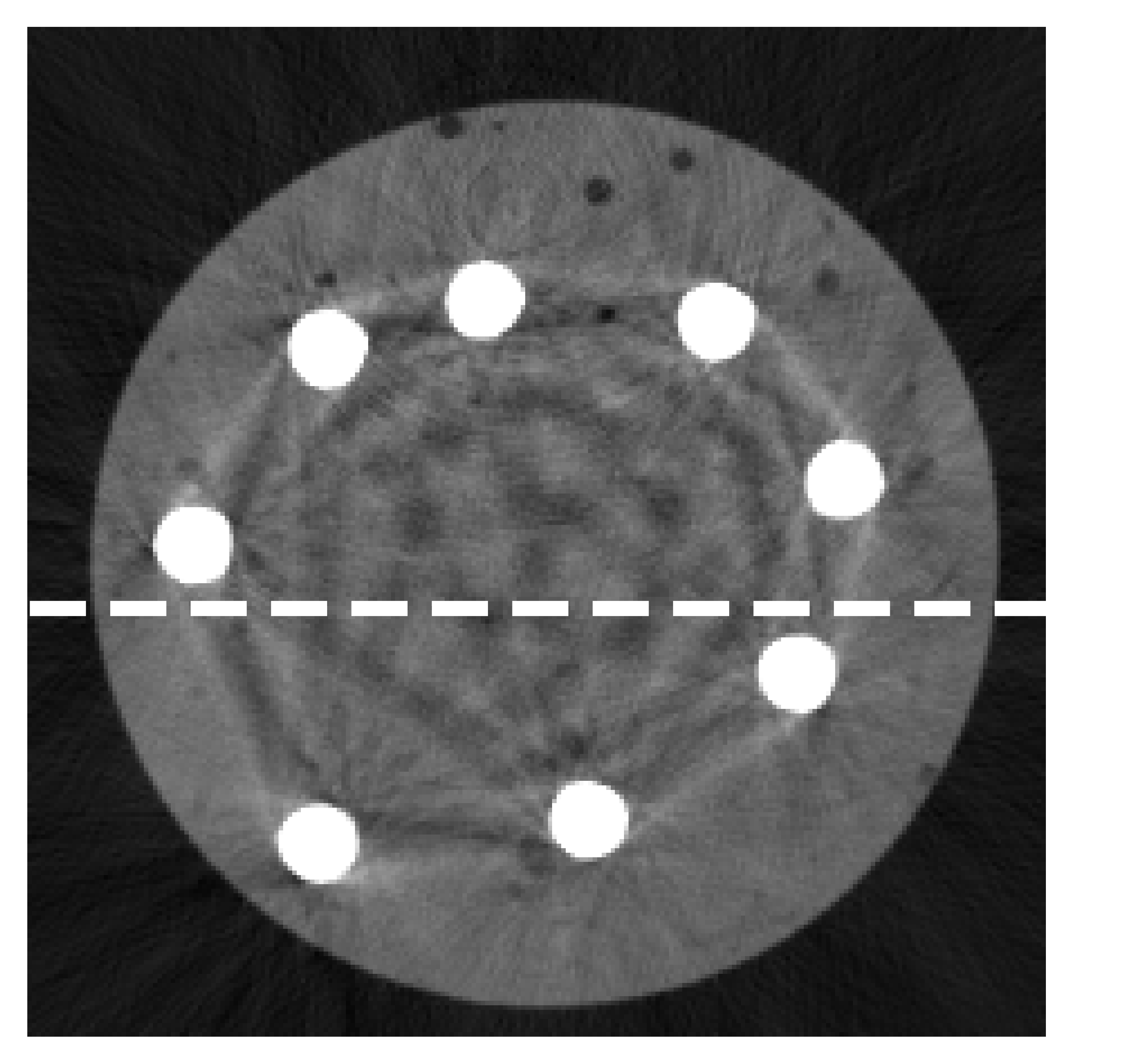}}{0.12in}{.1in}}\hspace{-0.8\baselineskip}
\subfloat{\topinset{\bfseries \textcolor{white}{(b)}}{\label{fig:CorrectlyEstimated}\includegraphics[width=0.42\linewidth]{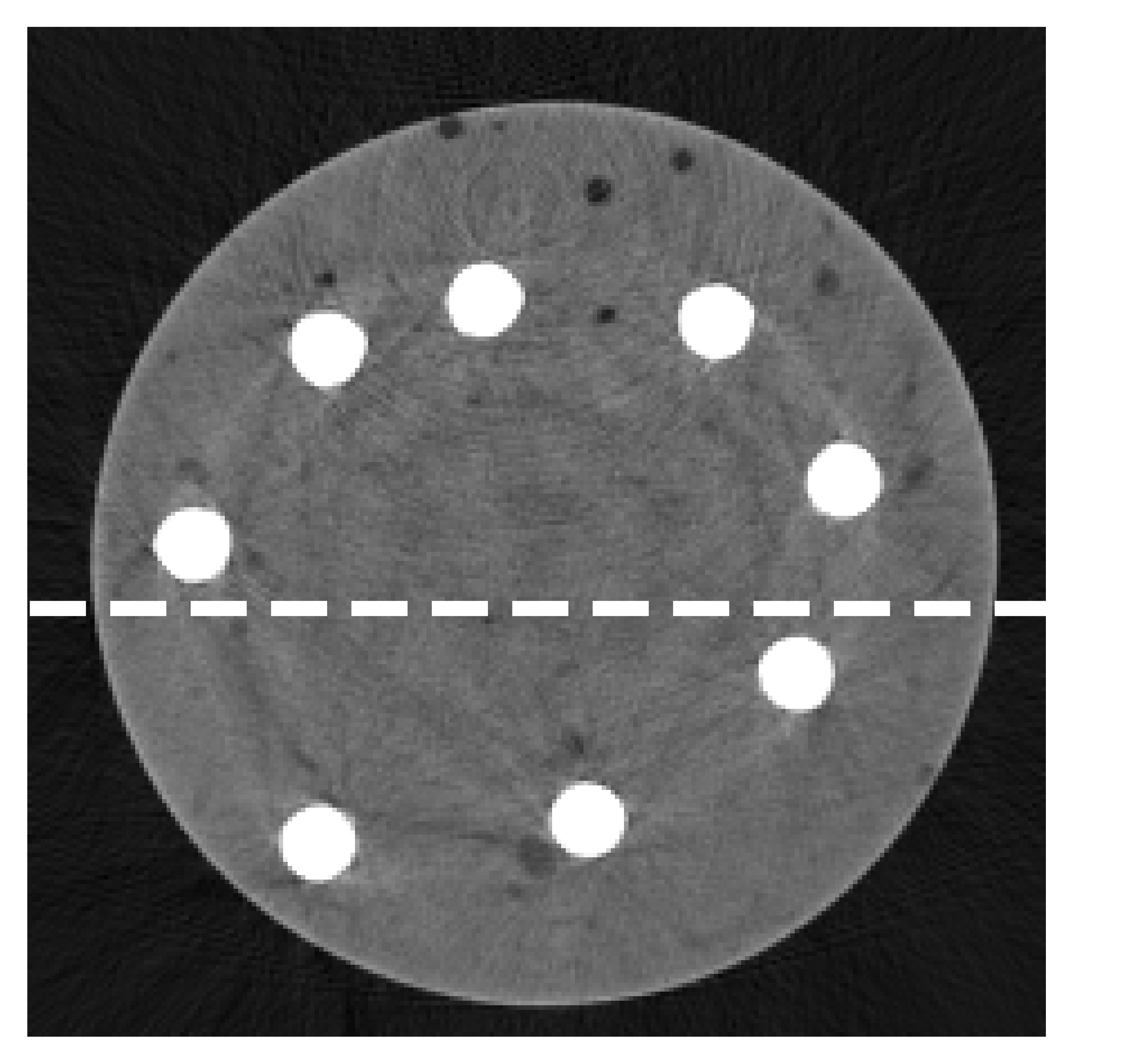}}{0.12in}{.1in}}\vspace{-1\baselineskip}
\subfloat{\topinset{\bfseries \textcolor{white}{(c)}}{\label{fig:LowerEstimated}\includegraphics[width=0.42\linewidth]{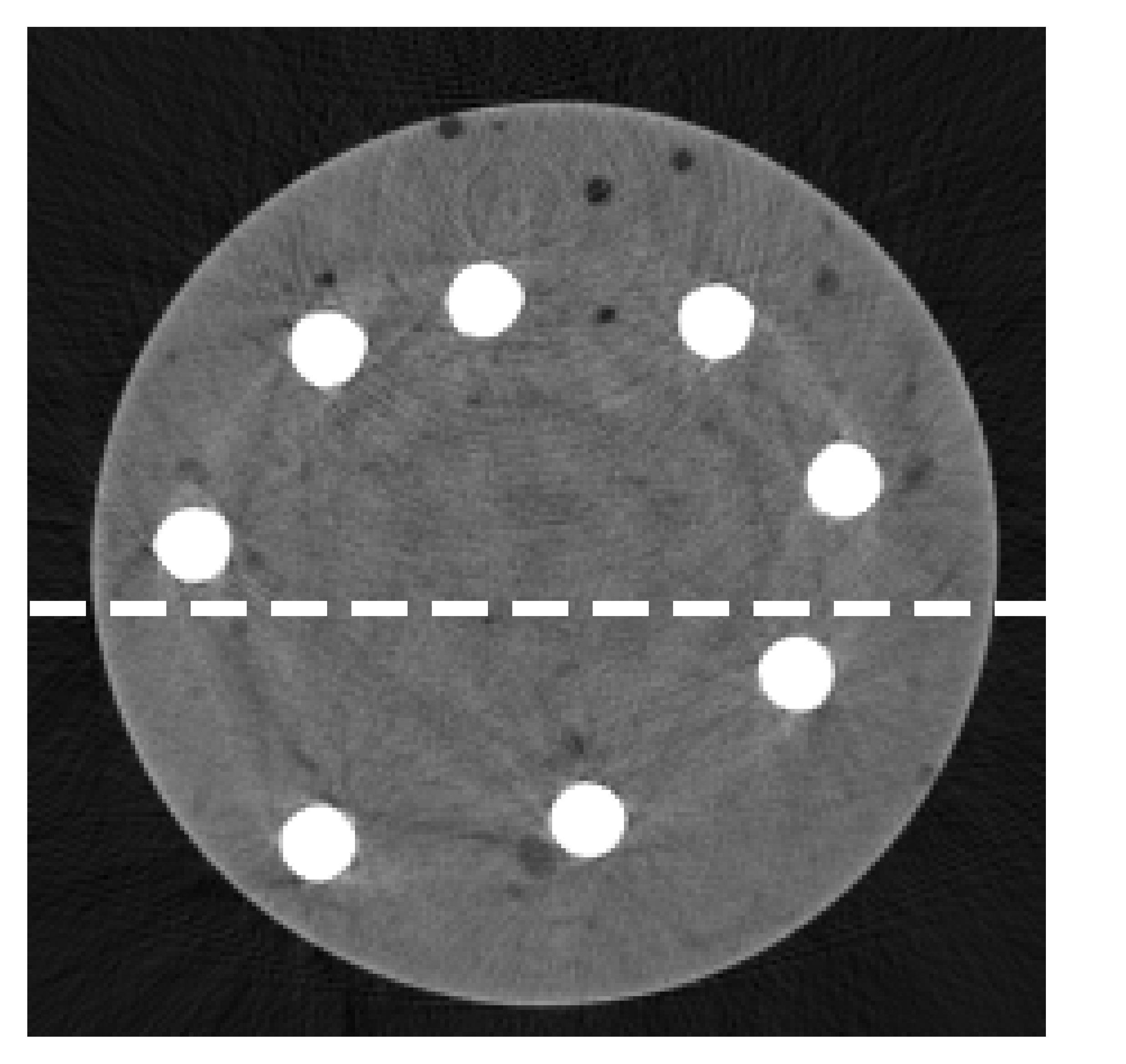}}{0.12in}{.1in}}\hspace{-0.8\baselineskip}
\subfloat{\topinset{\bfseries \textcolor{white}{(d)}}{\label{fig:UpperEstimated}\includegraphics[width=0.42\linewidth]{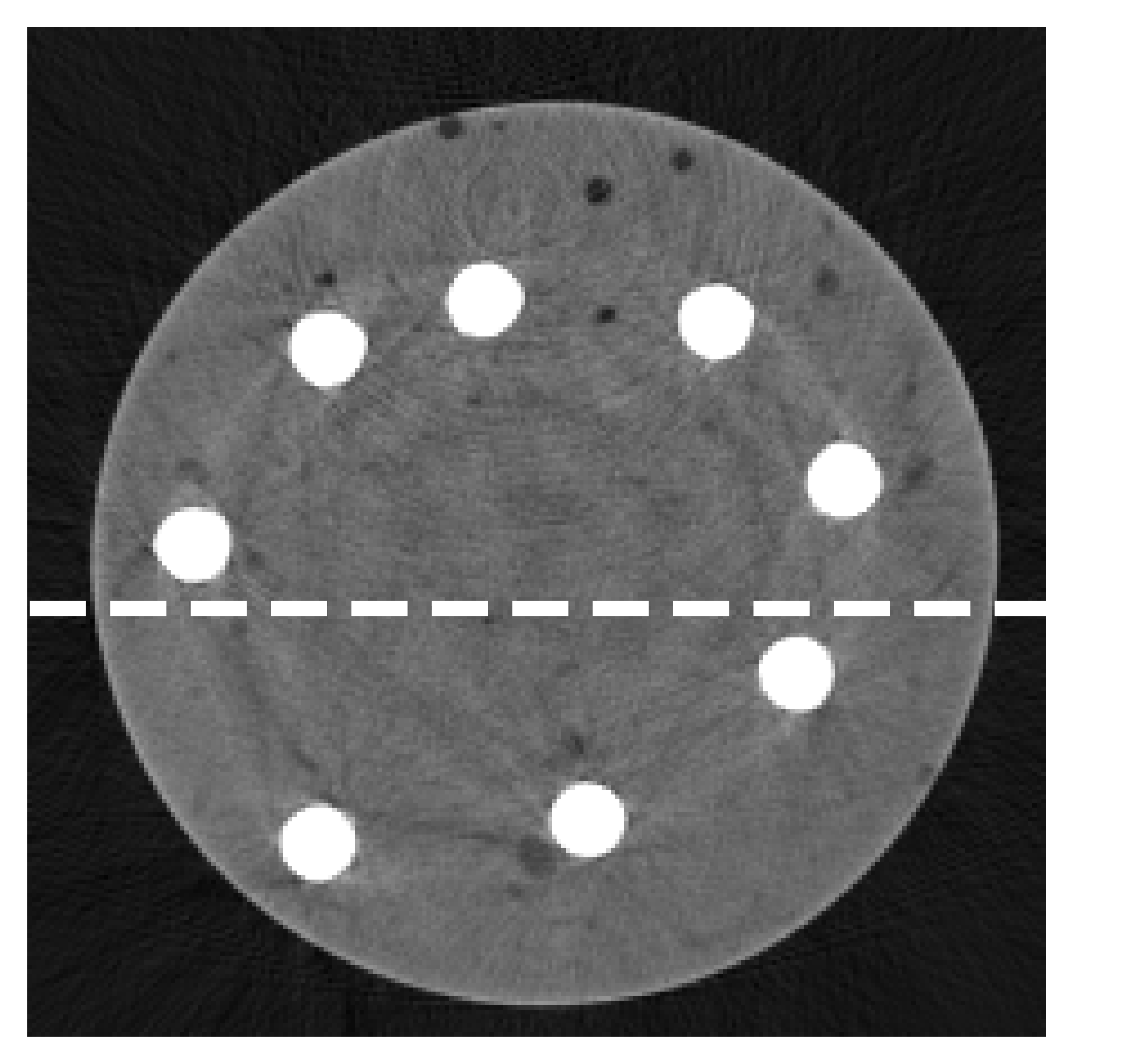}}{0.12in}{.1in}}\vspace{-0.1\baselineskip}
\subfloat{\topinset{\bfseries \textcolor{black}{(e)}}{\label{fig:depositenergy}\includegraphics[width=0.6\linewidth]{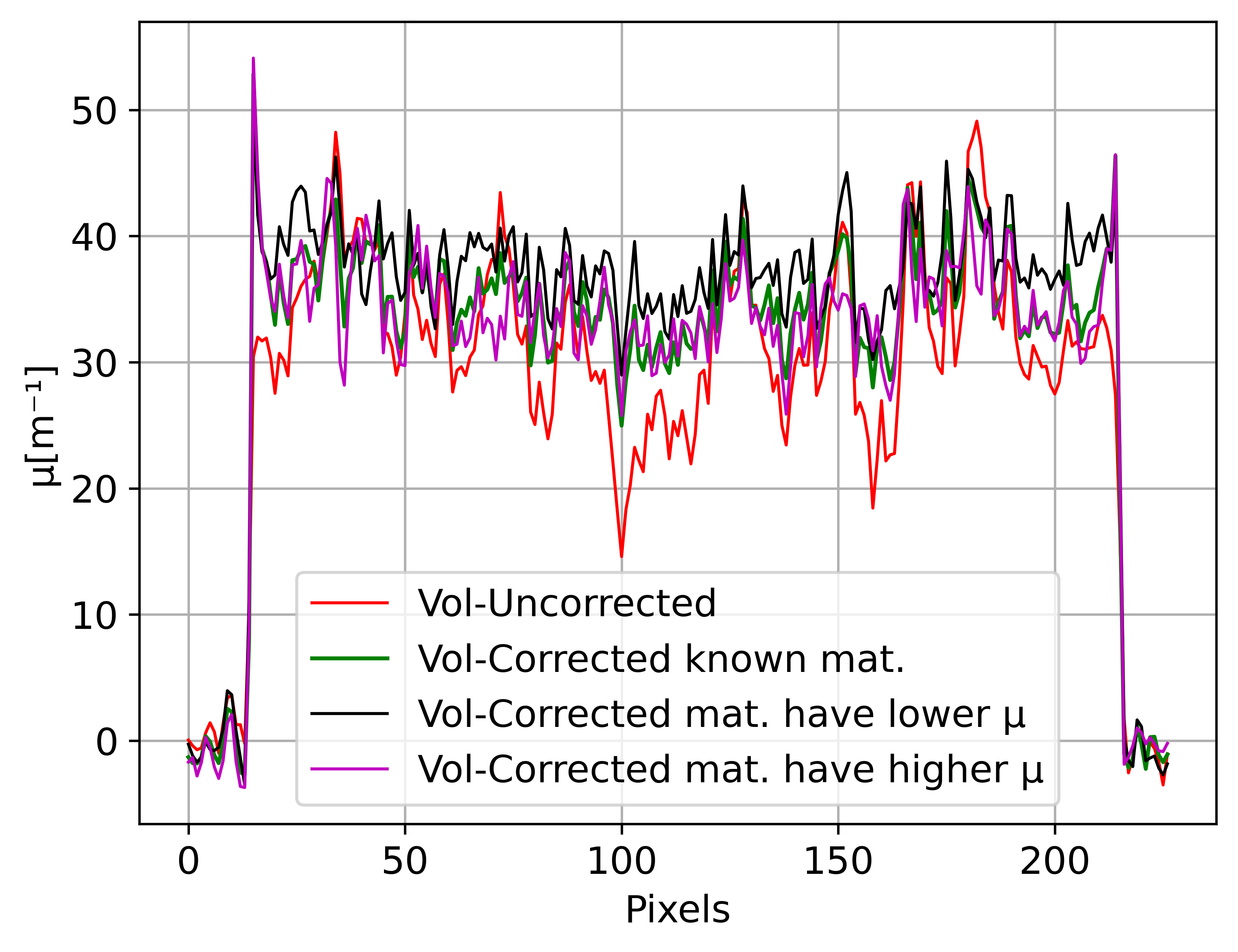}}{0.12in}{.27in}}\\

\caption{Second example which shows the results of the BH correction using the proposed method for different cases of materials estimation. (a) Slice from the uncorrected volume, (b) slice from the BH corrected volume assuming the materials are known, (c) slice from the BH corrected volume assuming the materials are wrongly estimated as silicon and titanium instead of the actual materials of cement and steel respectively, (d) slice from the BH corrected volume assuming the materials are wrongly estimated as aluminum and copper instead of the actual materials of the cement and steel respectively, (e) profile lines of the images in (a), (b), (c), and (d). The profiles are marked by white dashed lines in (a), (b), (c), and (d).}
\label{RobutOfMatEstim2}
\end{figure}

Fig. \ref{RobutOfMatEstim} and Fig. \ref{RobutOfMatEstim2} show the effect of the wrong estimation of the materials on the BH correction results for two different objects made from cement and steel. In these figures, three cases of materials estimation were tested. In the first case, the materials are assumed to be known. The results of the BH correction for the two objects in this case are shown in Fig. \ref{RobutOfMatEstim}\subref{fig:CorrectlyEstimated} and Fig. \ref{RobutOfMatEstim2}\subref{fig:CorrectlyEstimated}. In the second case, the materials from the uncorrected volume were assumed to have lower linear attenuation values than the values of the same materials on the look-up table. The deviation is assumed to be $20\%$ between the two values. In this case, the first material is selected by the material estimation approach as silicon and the second one is selected as titanium instead of cement and steel respectively. The BH correction results of this case are shown in Fig. \ref{RobutOfMatEstim}\subref{fig:LowerEstimated} and Fig. \ref{RobutOfMatEstim2}\subref{fig:LowerEstimated}. While in the third case, the materials are assumed to have higher attenuation values, i.e., $20\%$ deviation was assumed again. The first material in this case is selected as aluminum and the second material is selected as copper. Fig. \ref{RobutOfMatEstim}\subref{fig:UpperEstimated} and Fig. \ref{RobutOfMatEstim2}\subref{fig:UpperEstimated} show the BH correction results of the last case. From these figures, it is shown that the results of the BH correction for all the cases are almost the same and the wrong estimation of the materials has almost no effect on the quality of the correction using the proposed BH correction method. 

Table \ref{tab:materials} shows the actual estimation of the materials by the proposed materials estimation method for the five experimental objects shown in Fig.~\ref{Allexampleswithoutscatter}. This table shows the materials estimation results in case the volumes used in the BH correction are near scatter-free volumes and scatter-corrupted volumes. It is shown that the materials were wrongly estimated in these examples in the two cases, this is mainly due to the difference between the linear attenuation values extracted from the uncorrected volumes of these objects and the values of the relevant materials in the look-up table. However, the results of the BH correction, shown in Fig.~\ref{Allexampleswithoutscatter} and Fig.~\ref{Allexampleswithscatter} for the case of without and with the scatter effect respectively, show that the cupping and streak artifacts were highly suppressed. This is because the wrongly estimated materials are close to the original materials used in the objects, thus the polychromatic projection estimation method can compensate for the small deviation which results from this wrong estimation of the materials.



\begin{table}[ht]
\captionsetup{justification=centering}
\centering
\caption{MATERIALS ESTIMATION OF FIVE DIFFERENT EXPERIMENTAL EXAMPLES BY THE PROPOSED MATERIALS ESTIMATION METHOD IN CASE OF WITHOUT AND WITH THE SCATTER EFFECT.}
\begin{tabular}{ | C{4.0em} | C{1.9cm}| C{1.8cm} | C{1.5cm}|  } 
\hline
Objects& Actual Materials& Without scatter & With scatter\\
\hline
Object 1& Al, Fe & Concrete, Fe & Si, Ti\\ 
\hline
Object 2& Cement, Steel & Si,Fe & Mg, Ti\\  
\hline
Object 3& Cement, Steel & Si, Fe & Mg, Ti\\ 
\hline
Object 4& Cement, Stones & Mg, Mg & Mg, Mg\\ 
\hline
Object 5& Al & Al & Ti\\ 
\hline
\end{tabular}
\label{tab:materials}
\end{table}

\subsubsection{Robustness of the Polychromatic Projection Estimation}
\label{robofPoly}
\begin{figure}[ht!]
\centering
\def\stackalignment{l}
\subfloat{\topinset{\bfseries \textcolor{black}{(a)}}{\label{fig:depositenergy}\includegraphics[width=0.5\linewidth]{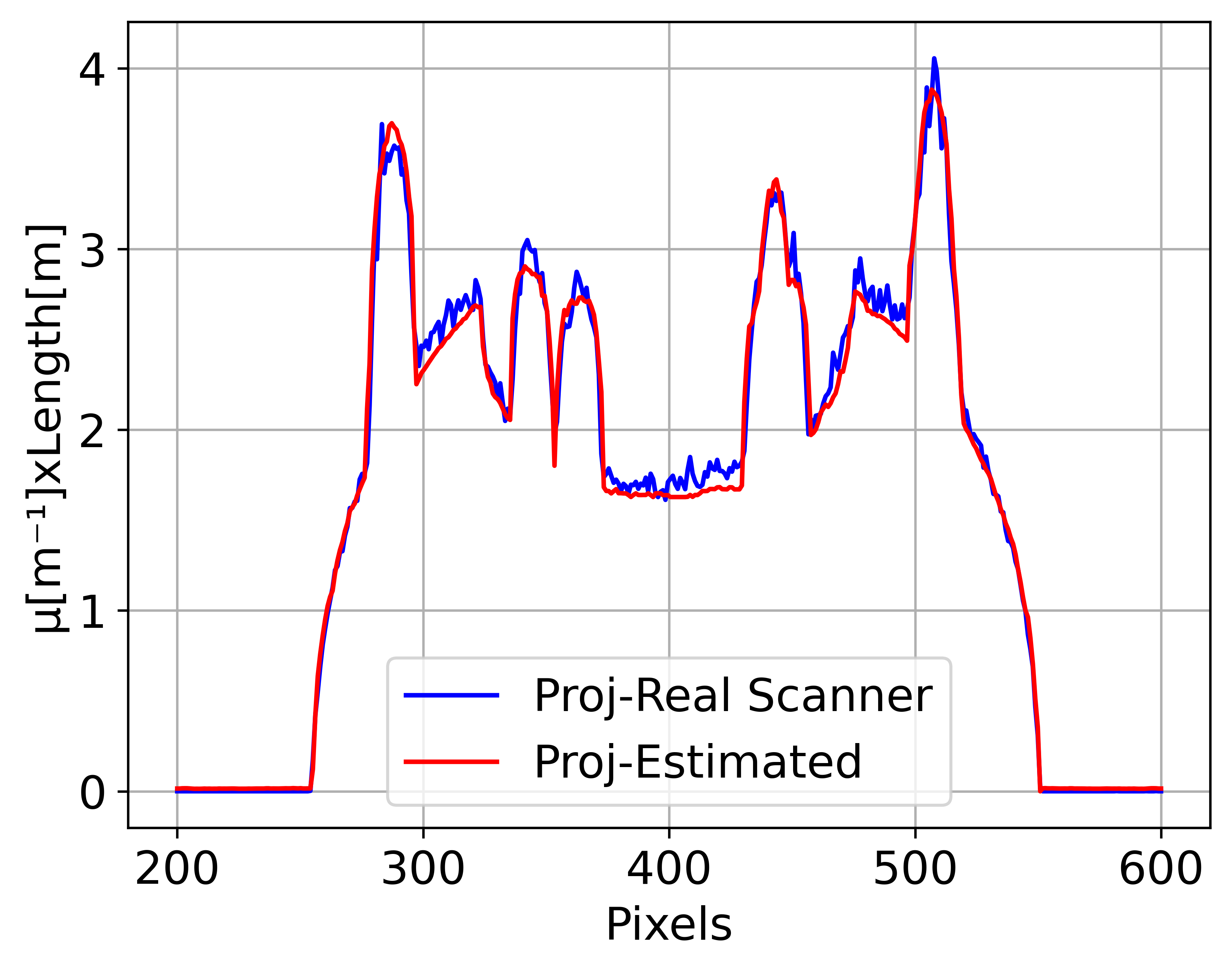}}{0.12in}{.27in}}
\subfloat{\topinset{\bfseries \textcolor{black}{(b)}}{\label{fig:detectorefficiency}\includegraphics[width=0.5\linewidth]{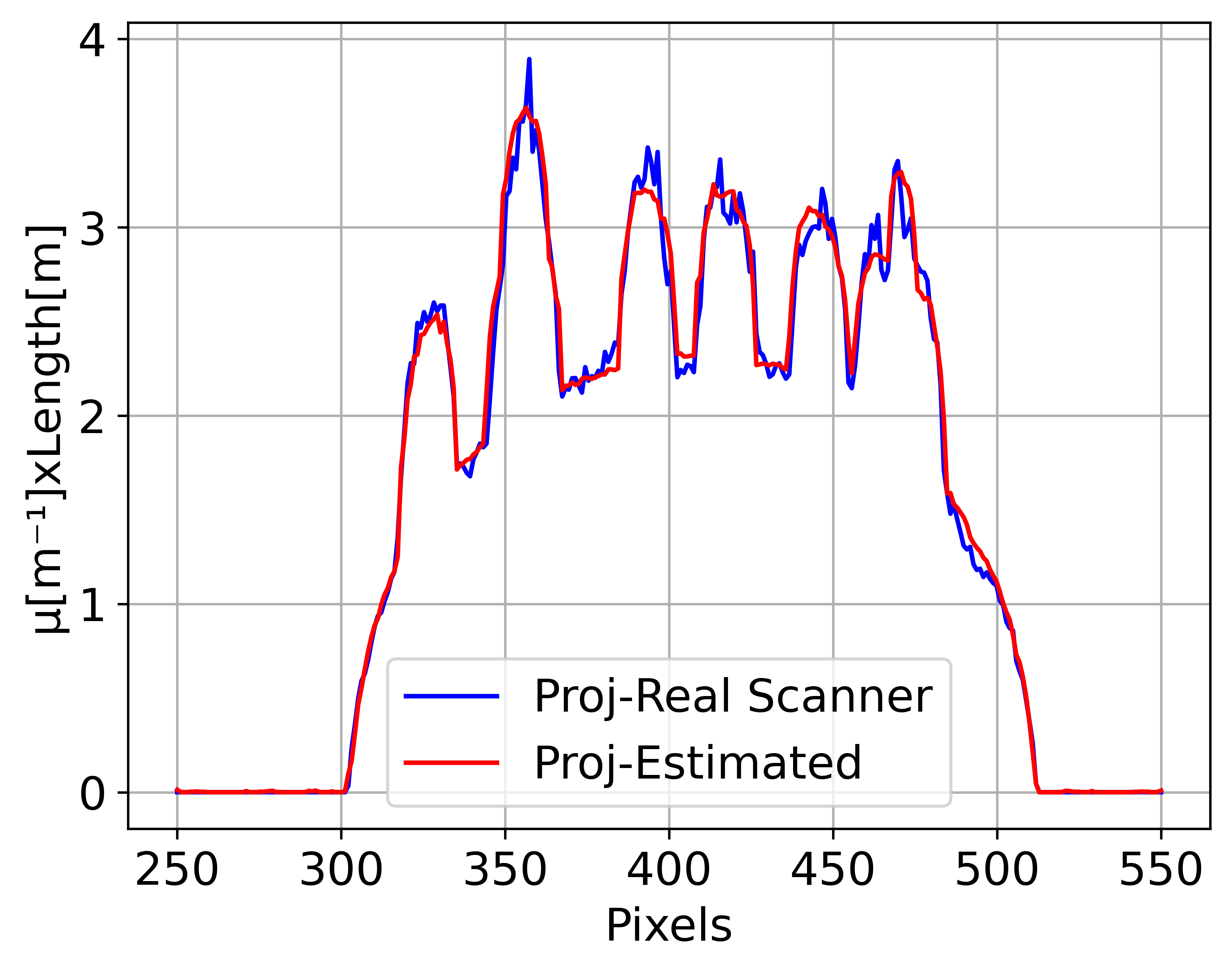}}{0.12in}{.27in}}\\[-0.1ex]

\caption{The central profile lines of the uncorrected projection from the scanner and the estimated polychromatic projection for two objects. (a) Comparison for the object shown in Fig. \ref{RobutOfMatEstim}, (b) comparison for the object shown in Fig. \ref{RobutOfMatEstim2}.}
\label{HowMatch}
\end{figure}

As stated before, to derive an accurate correction term that can correct the BH artifact of the projections from the scanner, the estimation of the polychromatic projection should be as close as possible to the original polychromatic projection from the scanner. Fig.~\ref{HowMatch} shows a comparison between the central profile lines of the projections from the scanner and the estimated polychromatic projections using the LSE method for the objects shown in Fig.~\ref{RobutOfMatEstim} and Fig.~\ref{RobutOfMatEstim2}. It is shown that the proposed LSE method can estimate a polychromatic projection which is very close to the original projection from the scanner for the two examples.

\subsubsection{Robustness of the Monochromatic Projection Estimation}
\label{robofMono}
\begin{figure}[ht!]
\centering
\def\stackalignment{l}

\subfloat{\topinset{\bfseries \textcolor{black}{(a)}}{\label{fig:depositenergy}\includegraphics[width=0.5\linewidth]{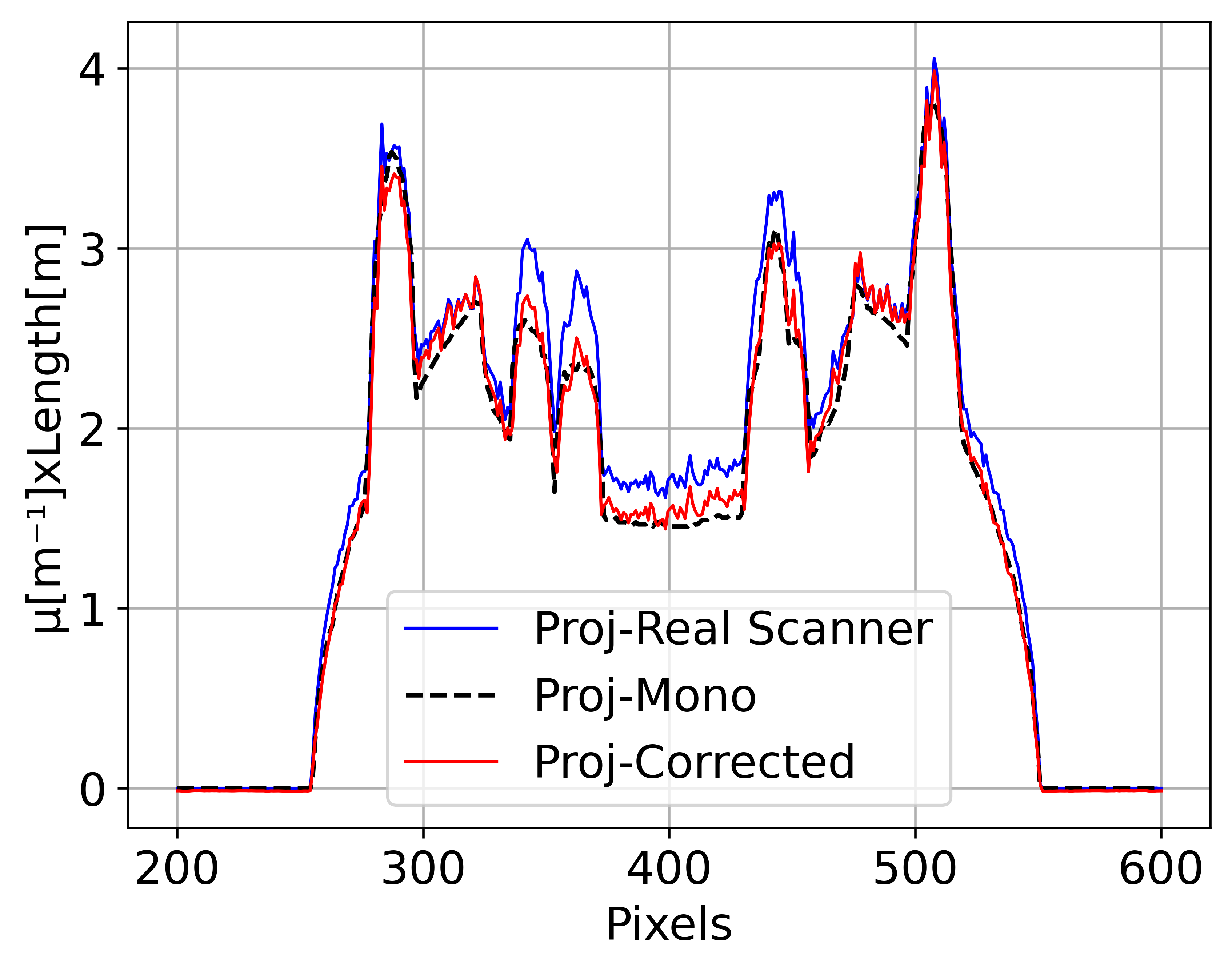}}{0.12in}{.27in}}
\subfloat{\topinset{\bfseries \textcolor{black}{(b)}}{\label{fig:detectorefficiency}\includegraphics[width=0.5\linewidth]{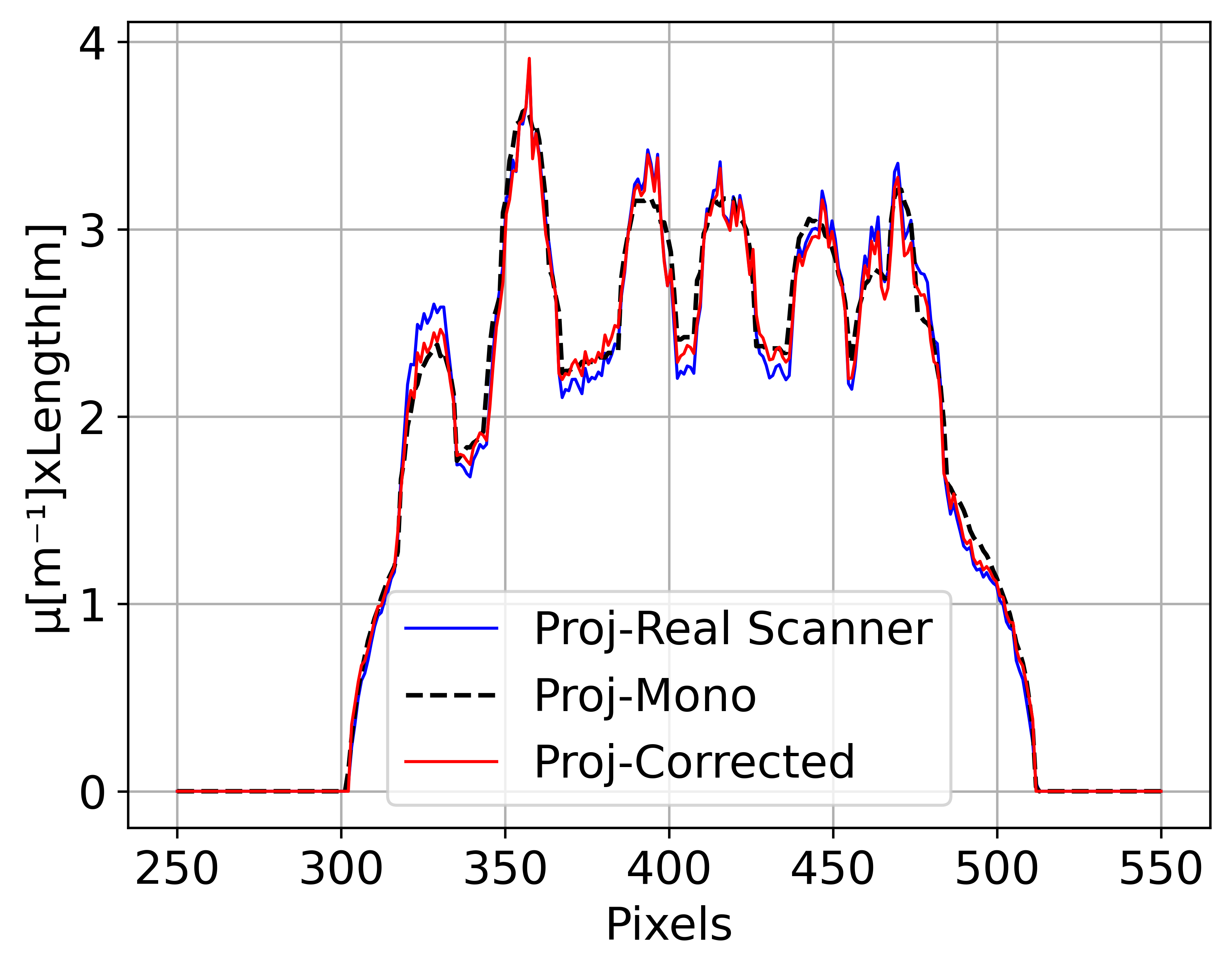}}{0.12in}{.27in}}\\[-0.1ex]

\caption{The central profile lines of the uncorrected projection from the scanner, the estimated monochromatic projection, and the BH corrected projection for two objects. (a) Comparison for the object shown in Fig. \ref{RobutOfMatEstim}, (b) comparison for the object shown in Fig. \ref{RobutOfMatEstim2}.}
\label{ProjsCorrected}
\end{figure}

The other part which influences the accuracy of the proposed BH correction algorithm is the ability to estimate a monochromatic projection which is equivalent to the polychromatic projection from the scanner. As mentioned before, effective monochromatic energy is obtained by iterating each energy bin separately. The energy bin which produces a projection that has the MMSE with the projection from the scanner is selected and used in the proposed BH correction algorithm. Fig.~\ref{ProjsCorrected} shows the center profile lines of the estimated monochromatic projections, the acquired projections and the corrected projections using the proposed BH correction algorithm for the objects shown in Fig.~\ref{RobutOfMatEstim} and Fig.~\ref{RobutOfMatEstim2}. It is shown that the estimated monochromatic projection is equivalent to the polychromatic projection from the scanner.

\subsubsection{Effect of Inaccurate Segmentation on BH Correction}
\label{roboagainstseg}
\begin{figure}[ht!]
\centering
\def\stackalignment{l}
\subfloat{\topinset{\bfseries \textcolor{white}{(a)}}{\label{fig:uncorrected}\includegraphics[width=0.33\linewidth]{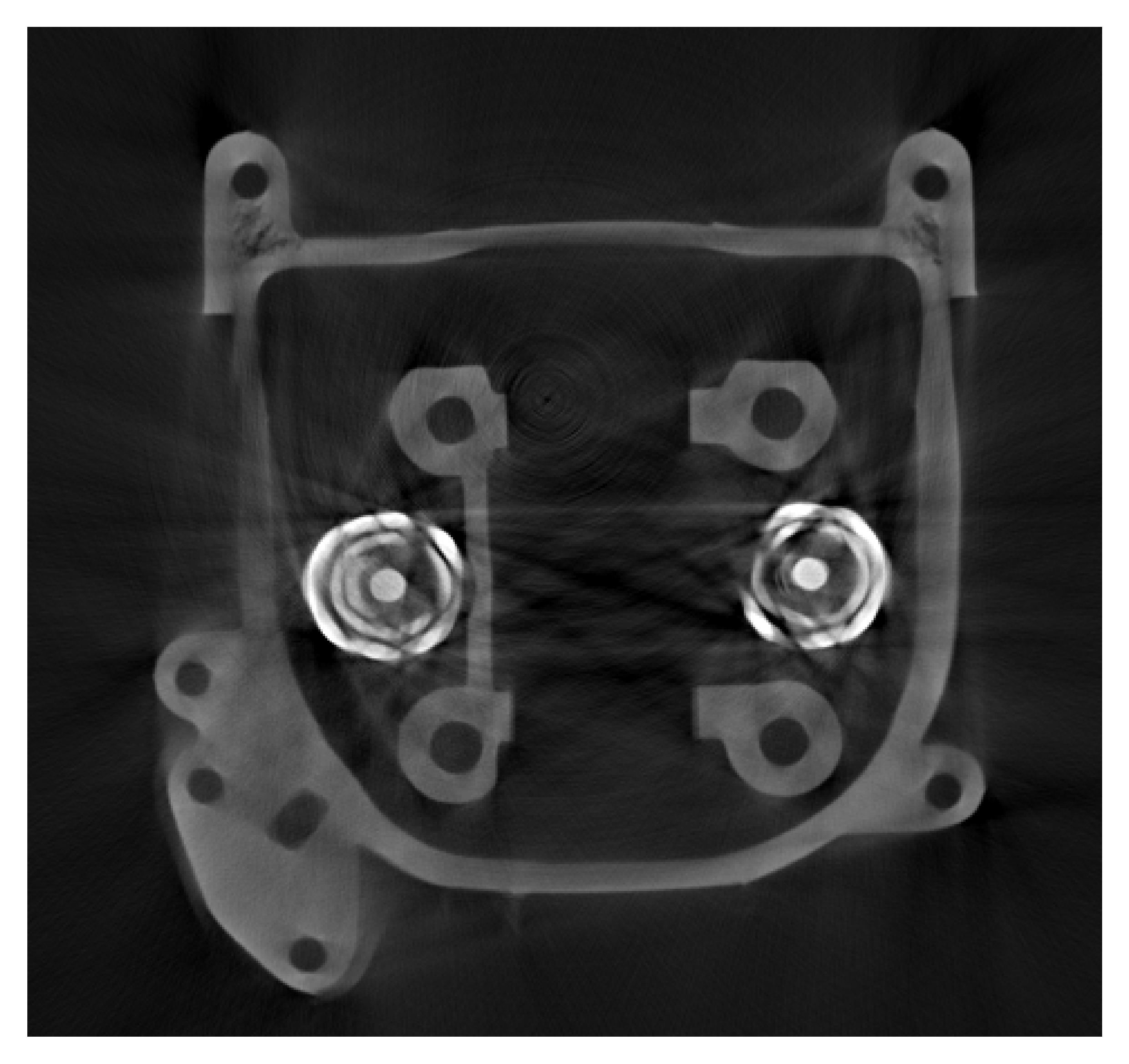}}{0.07in}{.05in}}
\subfloat{\topinset{\bfseries \textcolor{white}{(b)}}{\label{fig:segmentation}\includegraphics[width=0.33\linewidth]{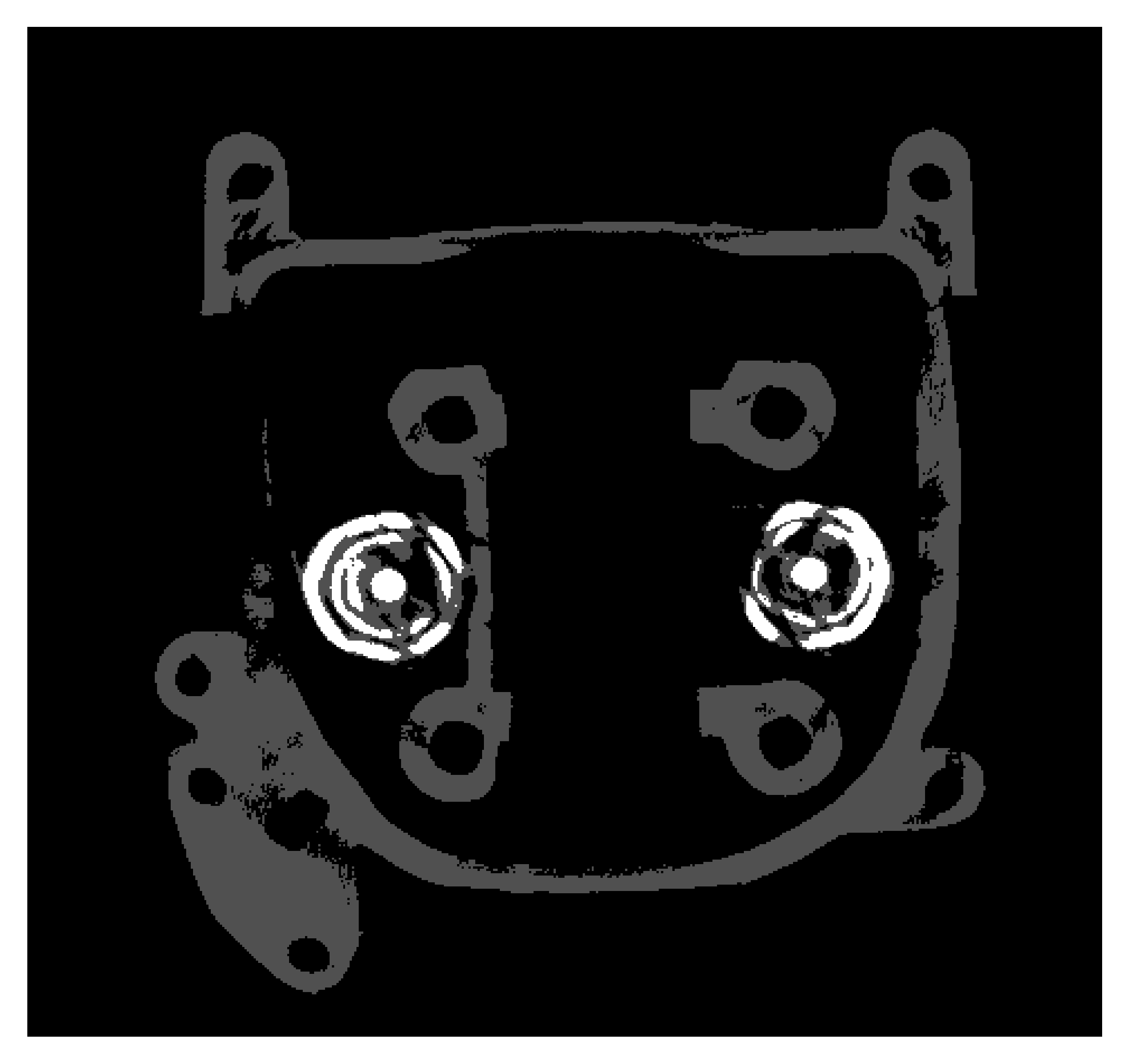}}{0.07in}{.05in}}
\subfloat{\topinset{\bfseries \textcolor{white}{(c)}}{\label{fig:corrected}\includegraphics[width=0.33\linewidth]{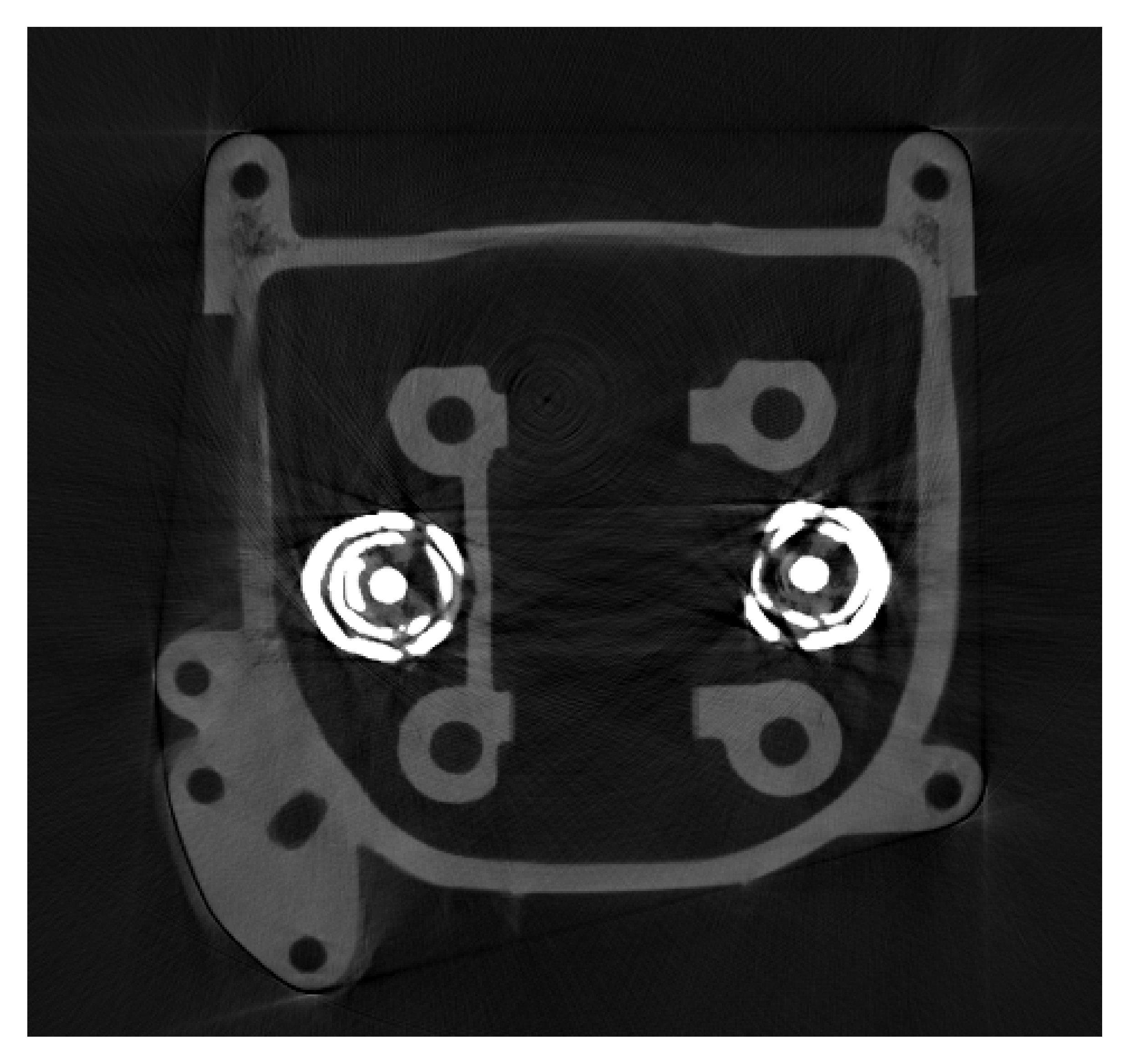}}{0.07in}{.05in}}\\[-0.1ex]

\caption{Segmentation effect on the BH correction results. (a) Slice from the uncorrected volume, (a) slice from the segmented uncorrected volume which is used in the proposed BH correction method to estimate the polychromatic and the monochromatic projections, (c) slice from the BH corrected volume.}
\label{SegmentationNotMatter}
\end{figure}

The estimated polychromatic and monochromatic projections are calculated based on the segmentation of the uncorrected volume contaminated by scattering and BH artifacts. Due to these artifacts, optimum segmentation of this volume is not always possible. Thus it is important to study the effect of the inaccurate segmentation on the BH correction result. Fig.~\ref{SegmentationNotMatter} shows the BH correction result using the proposed method from a near scatter-free volume acquired using a collimator but with BH artifacts. A slice from the uncorrected volume is shown in Fig.~\ref{SegmentationNotMatter}\subref{fig:uncorrected}. Due to the BH artifacts, some parts of this volume fall below the thresholds derived by the Otsu method and thus these parts were assigned to neither the aluminum part nor the steel part which are the two materials that this object contains. The result of the segmentation is shown in Fig.~\ref{SegmentationNotMatter}\subref{fig:segmentation}. Fig.~\ref{SegmentationNotMatter}\subref{fig:corrected} shows the result of the BH correction using the proposed method. It is shown that the missing parts in the segmented volume are still shown in the corrected one. This is mainly because the estimated polychromatic and the monochromatic projections were simulated from the segmented volume. Adding the correction term which is derived from these two estimated projections to the uncorrected projections will correct the relevant parts in the uncorrected volume leaving the rest of the volume which are not segmented untouched.   

\subsubsection{Effect of Scattering Artifacts on BH Correction}
\label{results}

\begin{figure}[ht!]
\centering
\def\stackalignment{l}

\subfloat{\topinset{\bfseries \textcolor{black}{(a)}}{\label{fig:centralscattercorrupted}\includegraphics[width=0.5\linewidth]{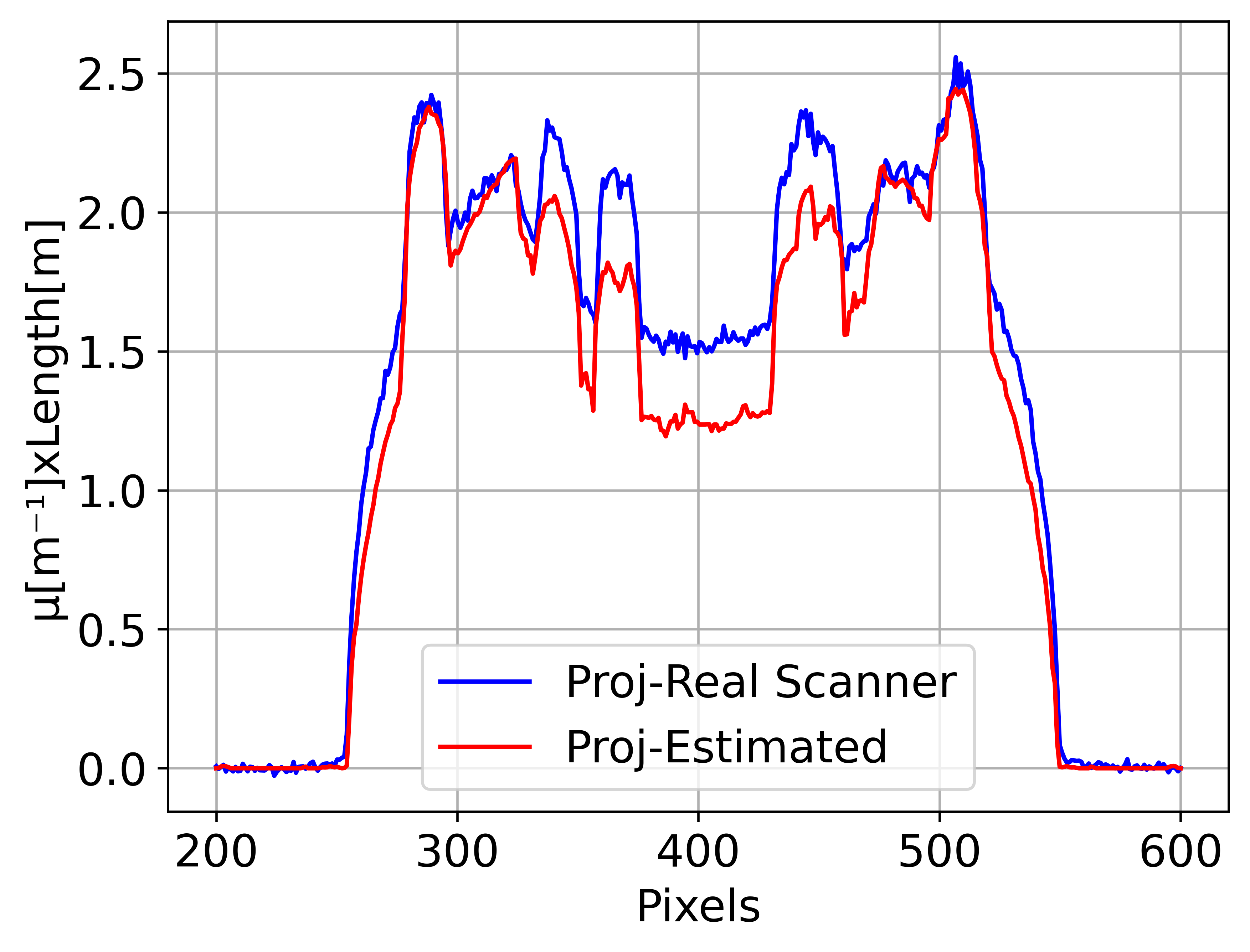}}{0.12in}{.27in}}
\subfloat{\topinset{\bfseries \textcolor{black}{(b)}}{\label{fig:centralscatterfree}\includegraphics[width=0.49\linewidth]{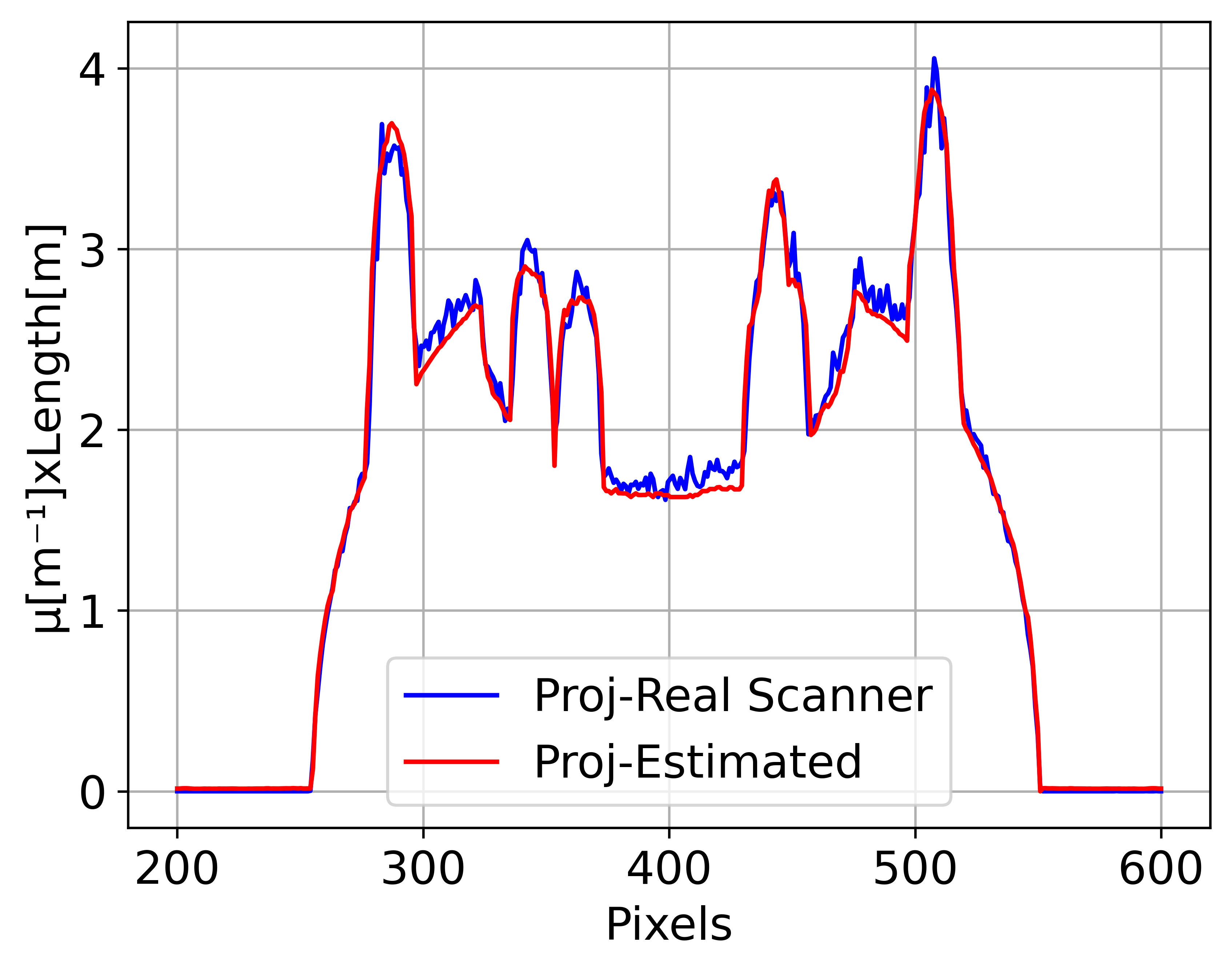}}{0.12in}{.27in}}\\[-0.1ex]

\caption{Scatter effect on the BH correction result. (a) The central profile lines of the scatter-corrupted projection from the scanner and the estimated polychromatic projection derived using the eight scatter-free simulated projections, (b) the central profile lines of the near scatter-free projection from the scanner acquired using a collimator and the estimated polychromatic projection derived using the eight scatter-free simulated projections.}
\label{Projsestimation with and without scatter effect}
\end{figure}

To obtain a good match between the projection from the scanner and the weighted sum of the eight simulated polychromatic projections from the LSE method, the simulated scan should mimic the actual scan. This means that if the acquired projections are corrupted by the scatter, the simulated polychromatic projections should also include the scatter. Although this is possible, as the model used to simulate the polychromatic projections is a MC photon forward projection model which is able to calculate the scatter accurately, this will prolong the time required by the BH correction method due to the long computation time required for the scatter calculation. Thus, the proposed BH correction method considers only the simulation of the projections using primary photons only and ignores the scatter. As a consequence, if the projections from the scanner are not scatter-free then the BH correction method is not able to estimate a polychromatic projection that matches completely the one from the scanner as shown in Fig.~\ref{Projsestimation with and without scatter effect}\subref{fig:centralscattercorrupted}. On the other hand, Fig.~\ref{Projsestimation with and without scatter effect}\subref{fig:centralscatterfree} shows that in case the projection from the scanner is near scatter-free, then the estimated polychromatic projection matches very well the one from the scanner.

\begin{figure}[ht!]
\centering
\def\stackalignment{l}

\subfloat{\topinset{\bfseries \textcolor{white}{(a)}}{\label{fig:centralscatterfree1}\includegraphics[width=0.34\linewidth]{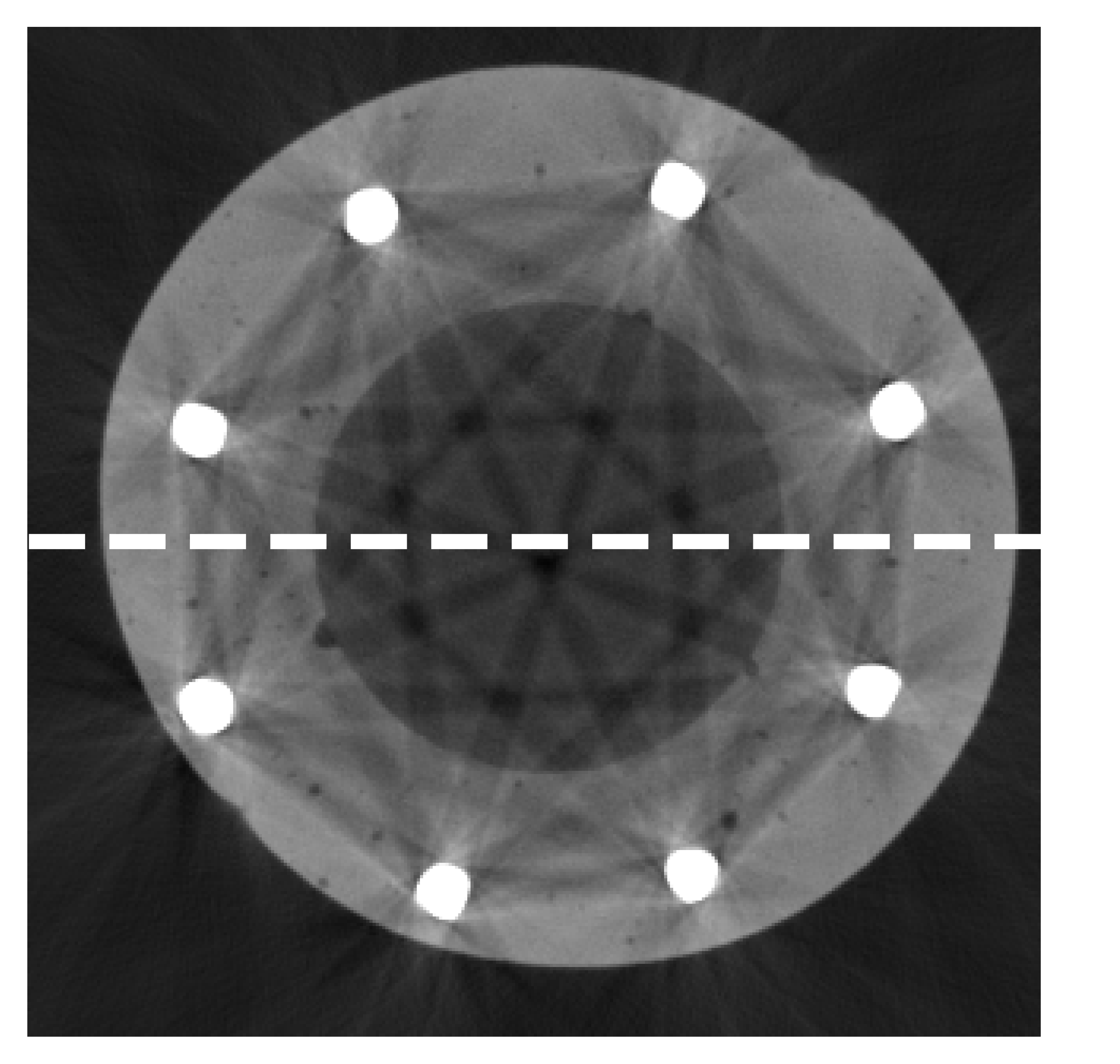}}{0.08in}{.08in}}\hspace{-0.7\baselineskip}
\subfloat{\topinset{\bfseries \textcolor{white}{(b)}}{\label{fig:ringCorrected_Collimator}\includegraphics[width=0.34\linewidth]{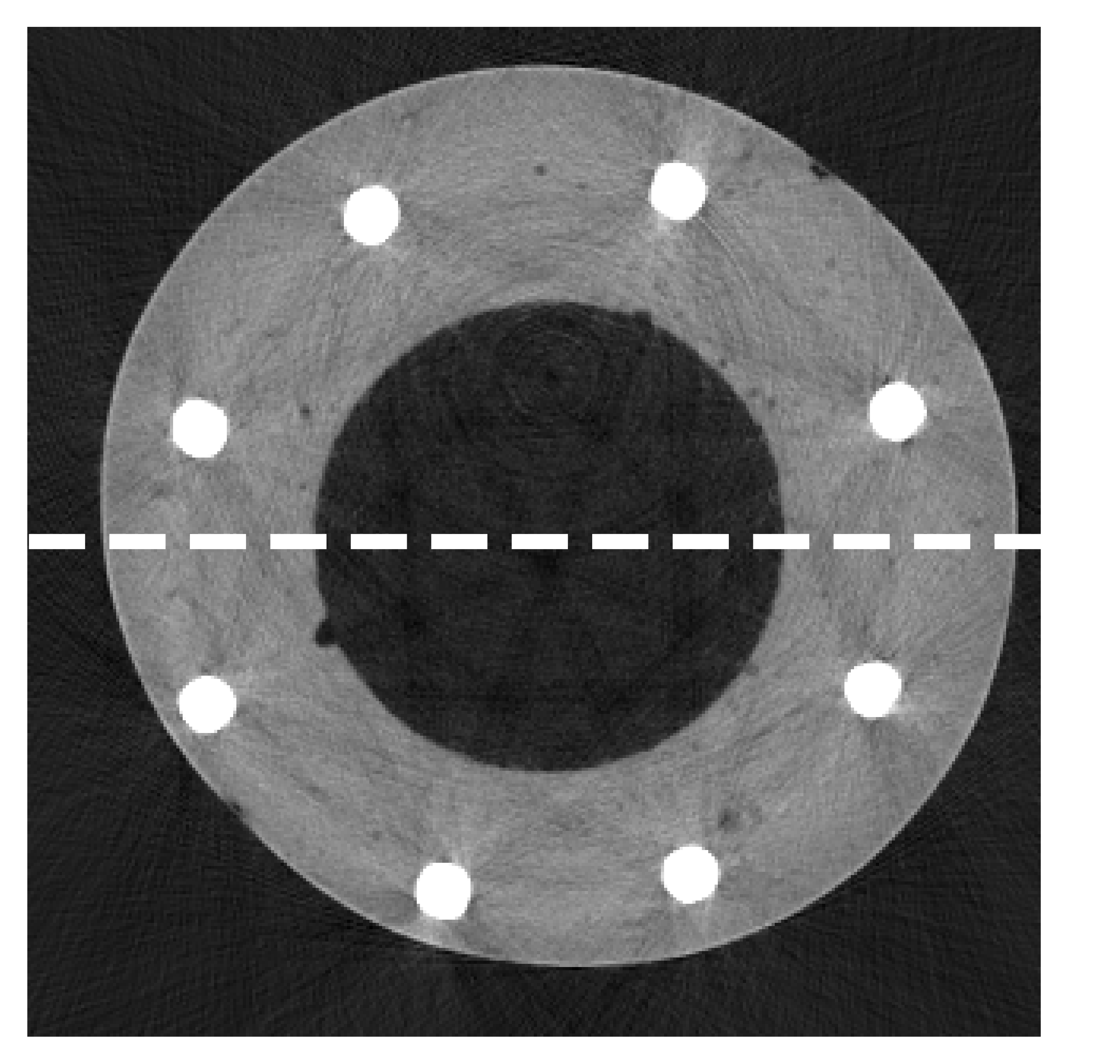}}{0.08in}{.08in}}\hspace{-0.7\baselineskip}
\subfloat{\topinset{\bfseries \textcolor{white}{(c)}}{\label{fig:ringCorrected_GPU}\includegraphics[width=0.34\linewidth]{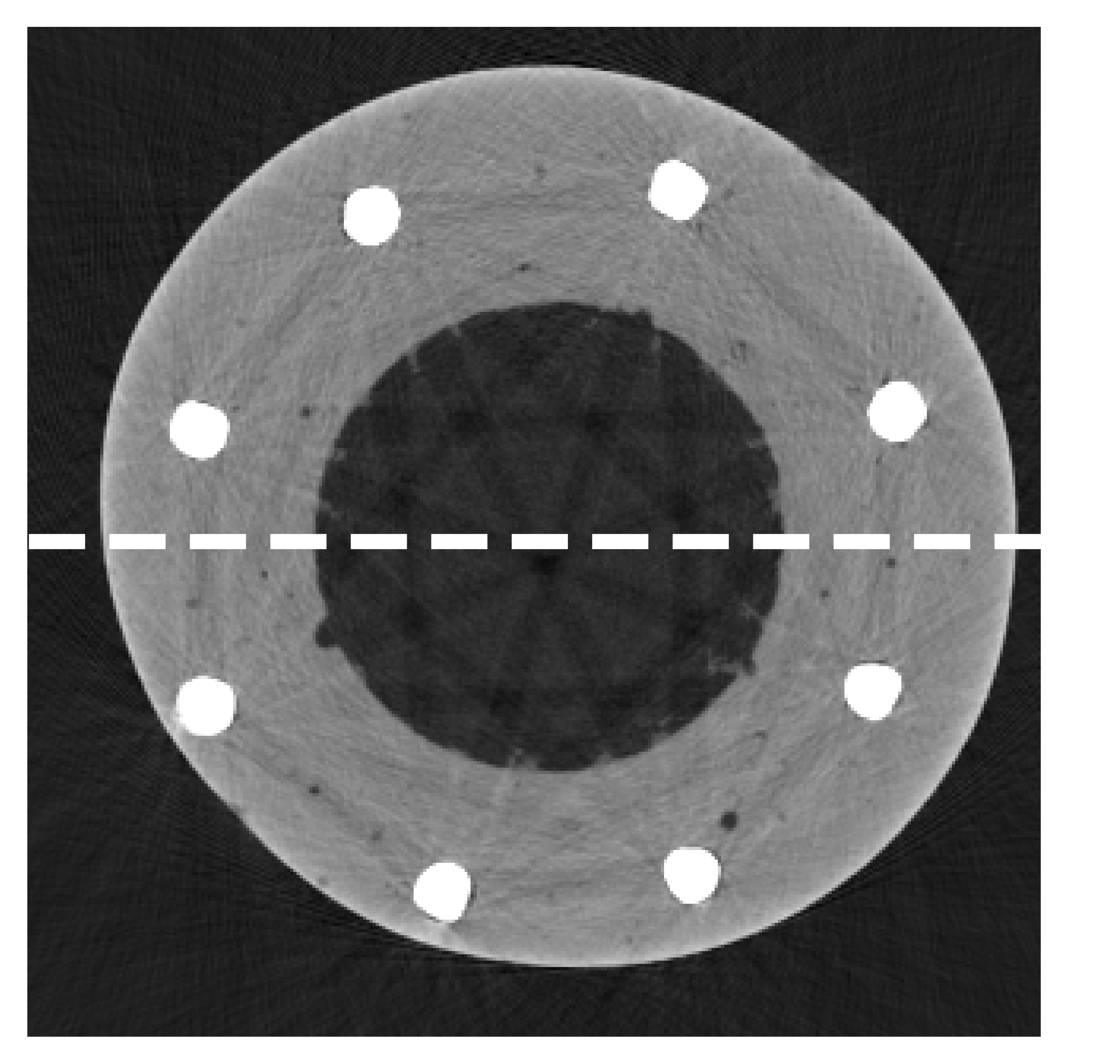}}{0.08in}{.08in}}\hspace{-0.7\baselineskip}\vspace{-1.07\baselineskip}

\subfloat{\topinset{\bfseries \textcolor{white}{(d)}}{\label{fig:centralscatterfree1}\includegraphics[width=0.34\linewidth]{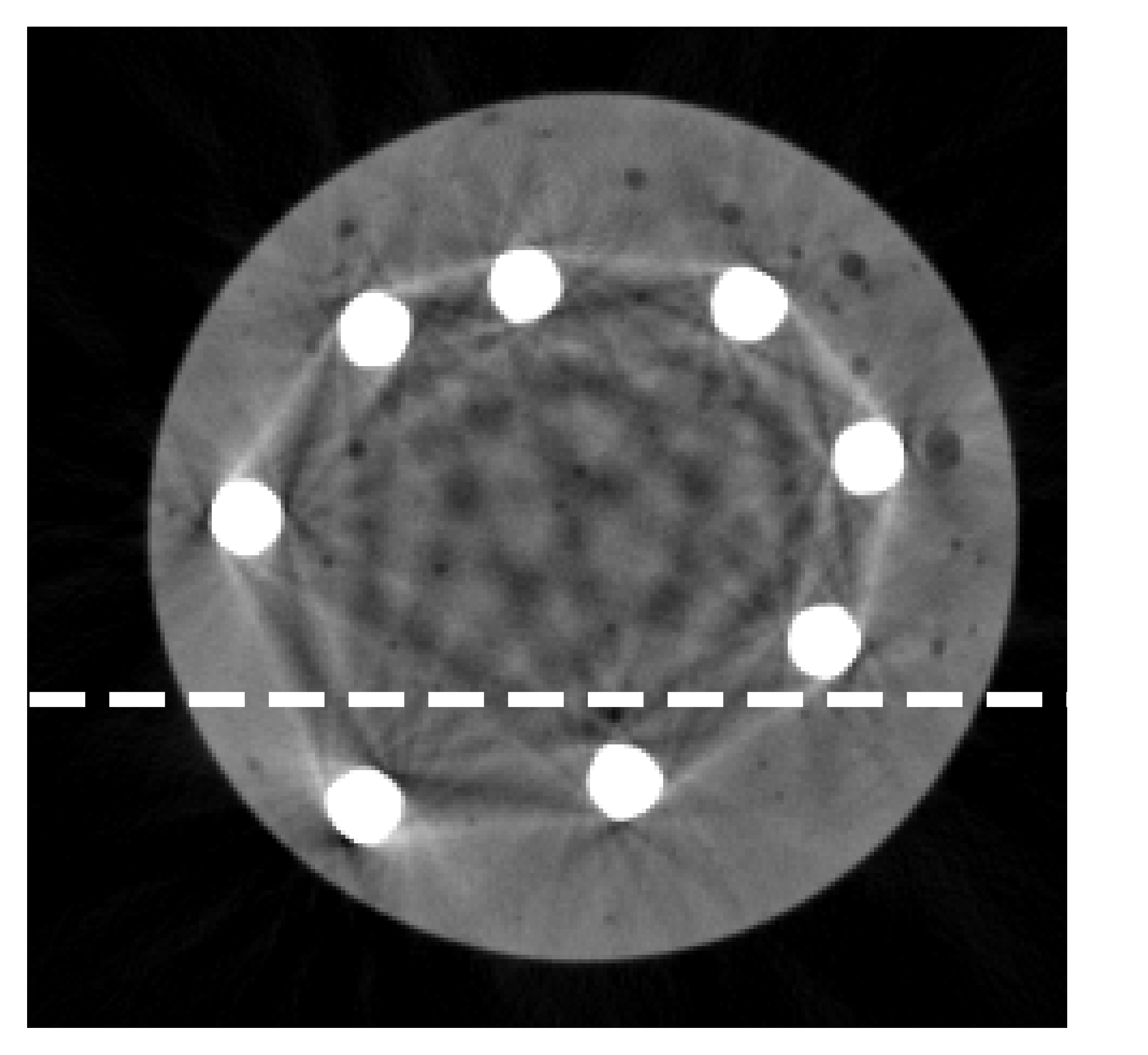}}{0.08in}{.08in}}\hspace{-0.7\baselineskip}
\subfloat{\topinset{\bfseries \textcolor{white}{(e)}}{\label{fig:CementSteel_Collimator}\includegraphics[width=0.34\linewidth]{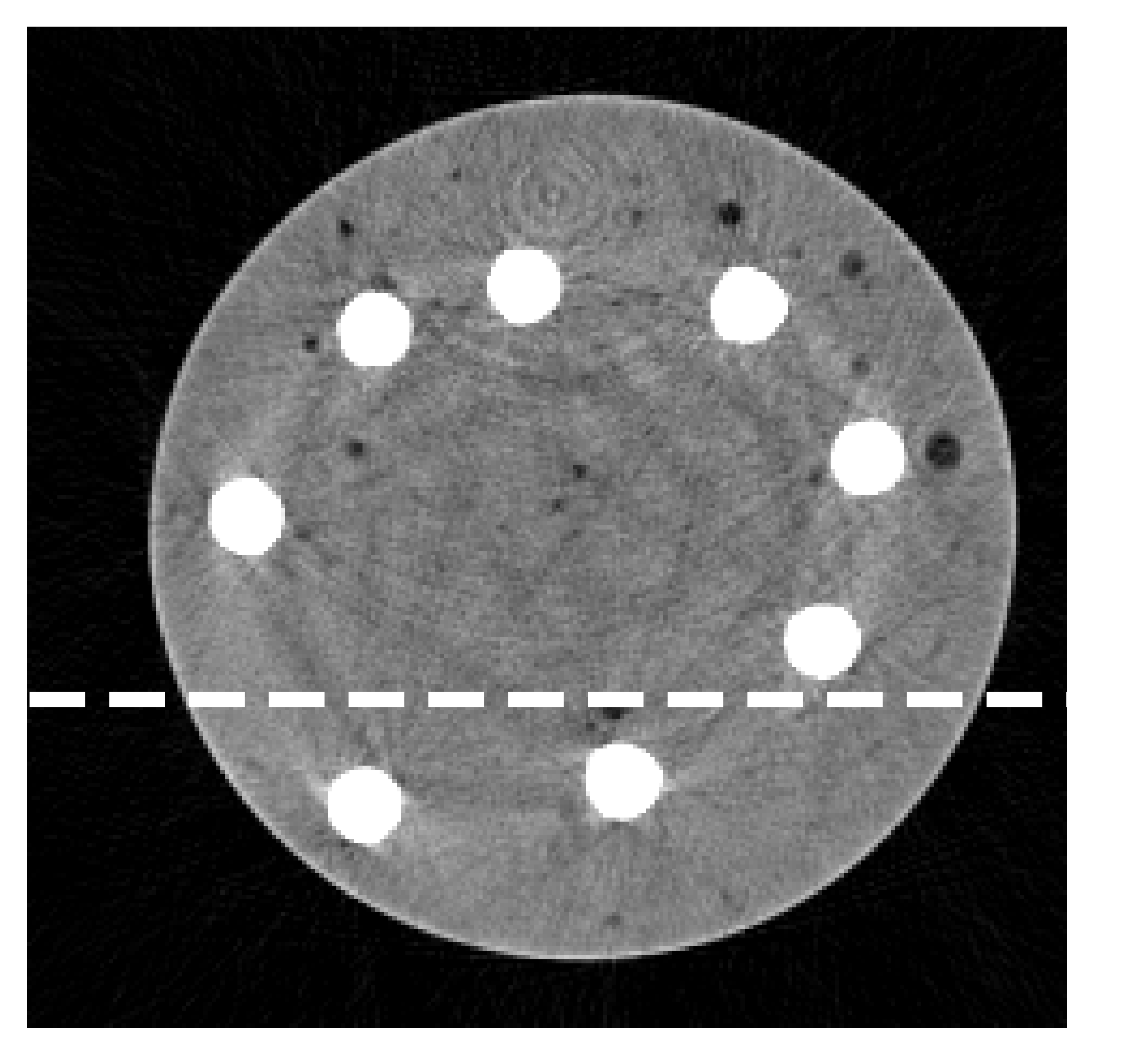}}{0.08in}{.08in}}\hspace{-0.7\baselineskip}
\subfloat{\topinset{\bfseries \textcolor{white}{(f)}}{\label{fig:CementSteel_GPU}\includegraphics[width=0.34\linewidth]{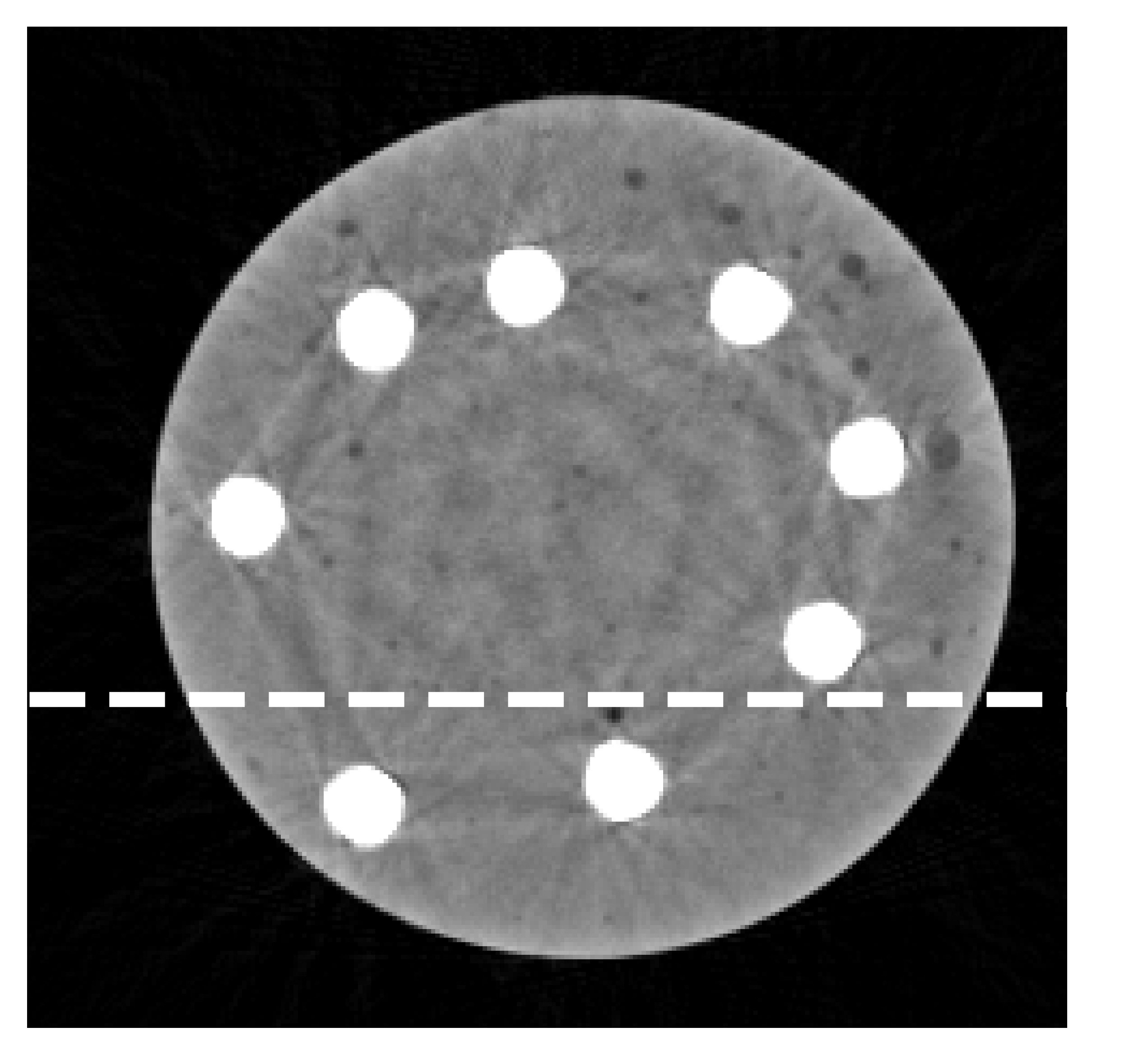}}{0.08in}{.08in}}\hspace{-0.7\baselineskip}\vspace{-0.8\baselineskip}

\subfloat{\topinset{\bfseries \textcolor{black}{(g)}}{\label{fig:centralscattercorrupted1}\includegraphics[width=0.49\linewidth]{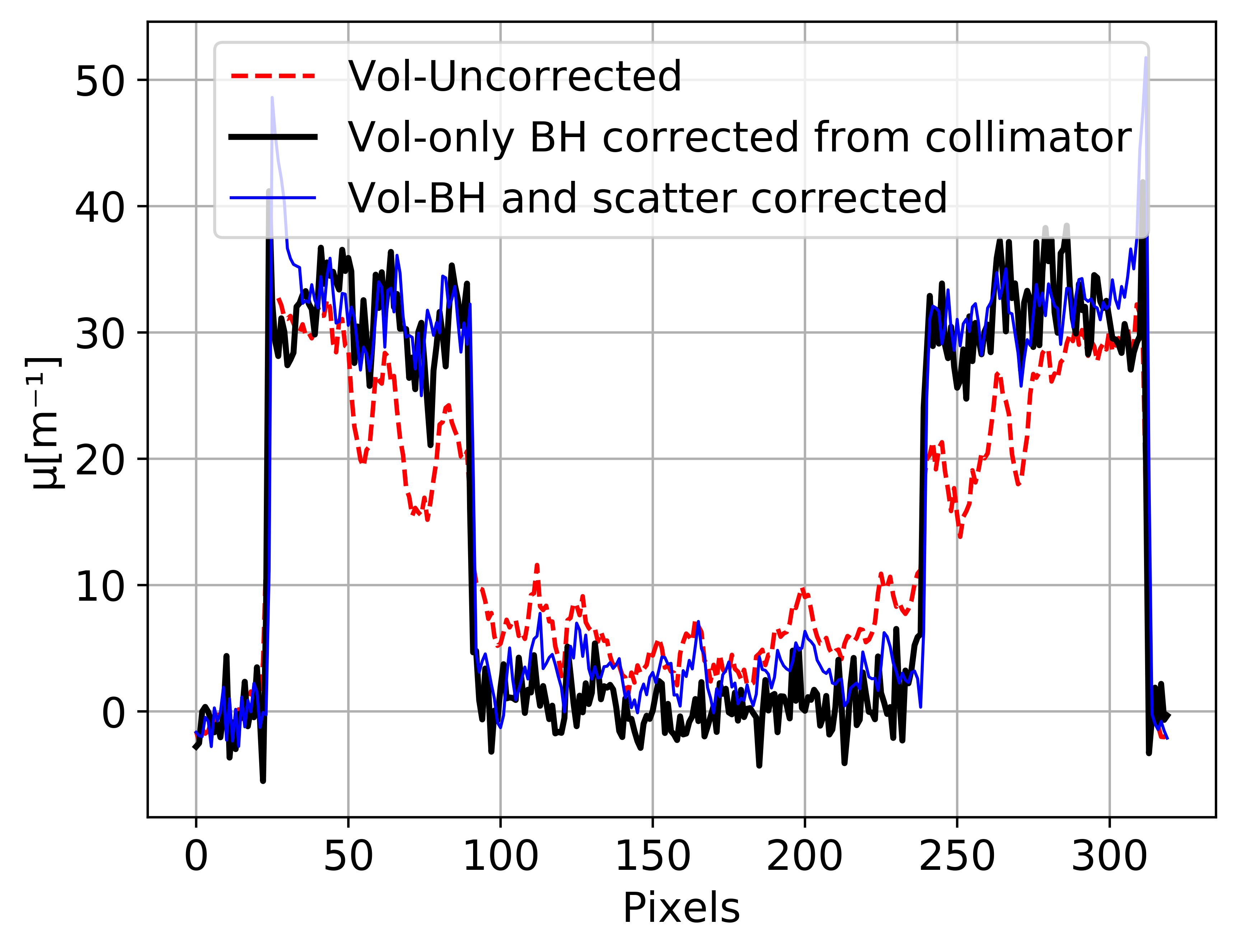}}{0.4in}{.22in}}
\subfloat{\topinset{\bfseries \textcolor{black}{(h)}}{\label{fig:centralscattercorrupted1}\includegraphics[width=0.5\linewidth]{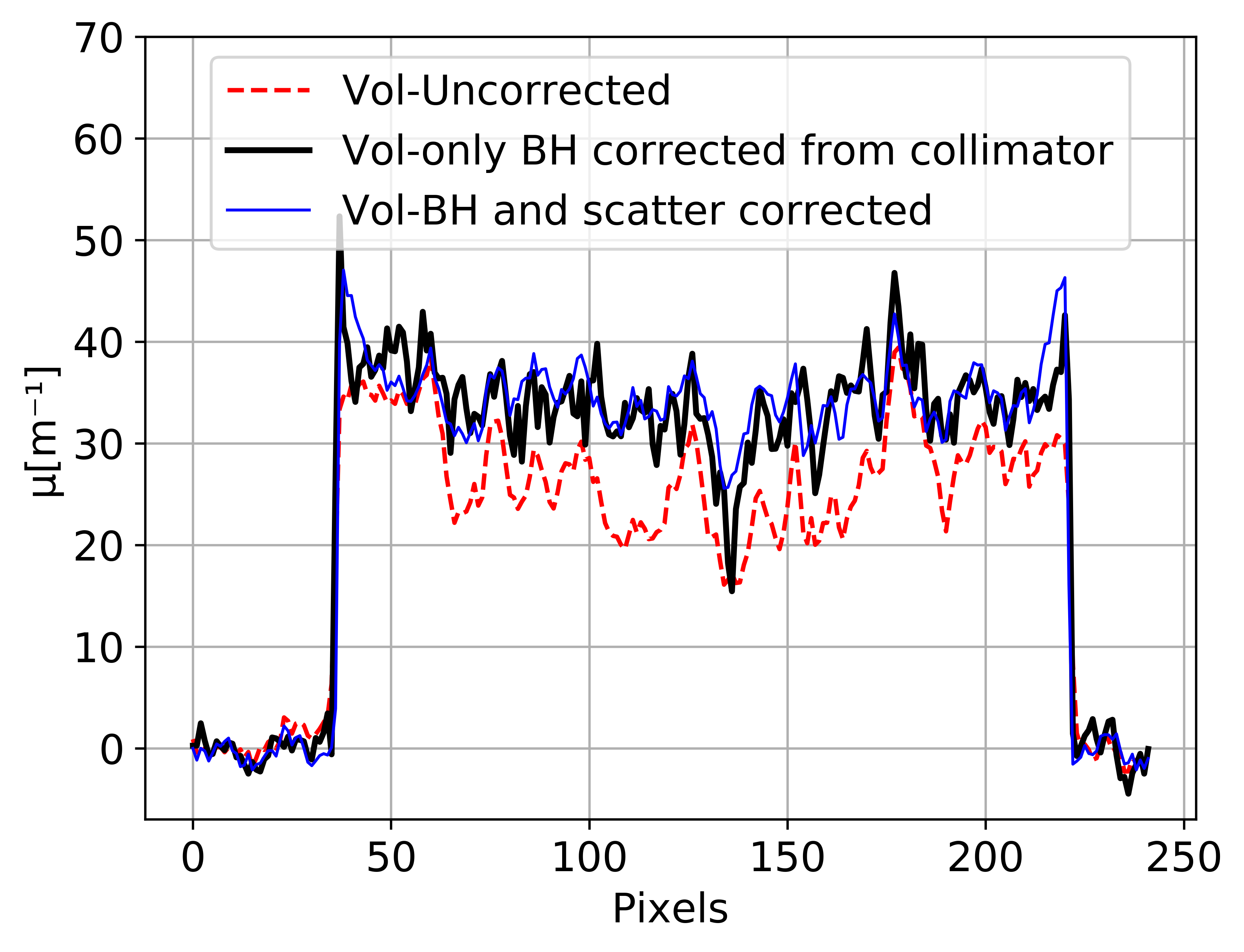}}{0.4in}{.22in}}

\caption{Comparison of the BH correction results when the BH correction is performed on near scatter-free volume acquired using the collimator and on scatter-corrected volume derived using the iterative scatter correction algorithm for two examples. (a) Uncorrected volume slice from the scanner of the cement and steel object 1, (b) slice from the BH corrected volume shows the result when the near scatter-free volume from the collimator is used (object 1), (c) slice from the BH corrected volume shows the result when the scatter-corrected volume is used (object 1), (d) Uncorrected volume slice from the scanner of the cement and steel object 2, (e) slice from the BH corrected volume shows the result when the near scatter-free volume from the collimator is used (object 2), (f) slice from the BH corrected volume shows the result when the scatter-corrected volume is used (object 2), (g) profile lines of (a), (b), and (c), (h) profile lines of (d), (e), and (f). The profiles are marked by white dashed lines in (a), (b), (c), (d), (e), and (f).}
\label{Projs from GPU and collimator}
\end{figure}

\begin{figure*}[ht!]
\centerline{\includegraphics[width=0.9\linewidth]{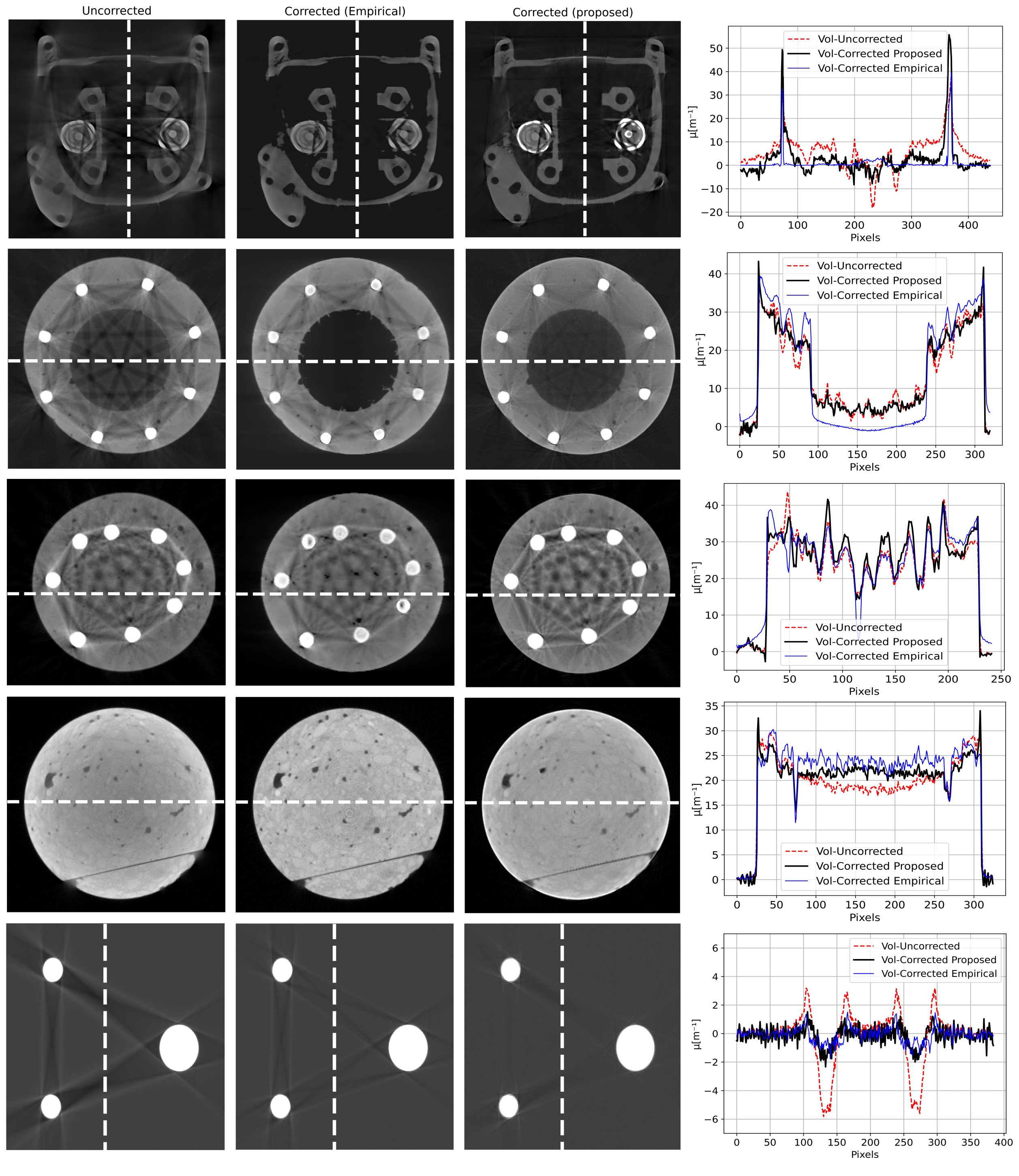}}
\caption{Experimental results of the proposed BH correction method using scatter-corrupted volumes from the scanner. The first column shows the slices from the uncorrected volumes. The second column shows the slices from the BH corrected volumes using the empirical method. The third column shows the slices from the BH corrected volumes using the proposed method. The last column shows the comparisons between the profile lines of the slices in columns 1 to 3. The object in the first row is made from aluminum and steel. The objects in the second and the third rows are made from cement and steel. The fourth object is a concrete cylinder. The last object represents three aluminum cylinders positioned away from each other in the space. The simulated projections used in the proposed BH correction are from primary photons only without scatter. The profiles are marked by white dashed lines in these sub-figures.}
\label{Allexampleswithscatter}
\end{figure*}

Several methods could be employed in order to get a near scatter-free result from the scanner which can be used in the BH correction algorithm. A simple collimator of two copper blocks of a $2\,cm\times2\,cm\times4\,cm$ dimension has been placed in front of the X-ray source in which the long side was positioned perpendicularly with the exit window of the X-ray source. This suppresses the scatter and produces near scatter-free results since the collimator converts the cone-beam CT into a fan-beam that produces near scatter-free results. In addition, a powerful scatter correction method could also be used in order to get scatter-free results. In \cite{Alsaffar} an iterative scatter correction algorithm and a multi-GPU MC photon forward projection model have been implemented. The acquired images can be first corrected by scatter suppression using the iterative algorithm, then the BH artifacts in these images are corrected using the proposed method in this work. Especially, the scatter existing in the acquired image is iteratively corrected by using scatter estimated from MC simulation. In each iteration, the estimated scatter is subtracted from the acquired projections in order to derive a better scatter-corrected image. It is shown in \cite{Alsaffar} that near scatter-free images can be obtained using one to three iterations with MC scatter computation time less than the time required for the standard FBP reconstruction algorithm being the fastest and the most widely used reconstruction technique. Fig.~\ref{Projs from GPU and collimator} shows the BH correction results of two examples. In each example, the BH correction was performed after the removal of the scatter. Fig.~\ref{Projs from GPU and collimator}\subref{fig:ringCorrected_Collimator} and Fig.~\ref{Projs from GPU and collimator}\subref{fig:CementSteel_Collimator} show the BH correction results of the two examples using near scatter-free volumes from the collimator. While Fig.~\ref{Projs from GPU and collimator}\subref{fig:ringCorrected_GPU} and Fig.~\ref{Projs from GPU and collimator}\subref{fig:CementSteel_GPU} show the BH correction results of the same examples using scatter-corrected volumes from the aforementioned method. It is shown that the result of the BH correction for the near scatter-free volume from the collimator match very well the result of the BH correction performed using the scatter-corrected volume. Fig.~\ref{Allexampleswithscatter} shows the experimental results of the proposed BH correction method for the same five experimental objects shown in Fig.~\ref{Allexampleswithoutscatter} but all the volumes, in this case, are corrupted by the scatter in addition to the BH artifacts. Although the results of the BH correction in Fig.~\ref{Allexampleswithscatter} show good improvement in the presence of scattering, the results in Fig.~\ref{Allexampleswithoutscatter} show that the proposed method performs better and produces superior BH correction results in case the correction was performed using near scatter-free volumes from the scanner.

\section{Conclusion}

In this work, a blind and robust BH correction method with unknown materials is proposed. The materials within the uncorrected volume are first estimated using a look-up table of the measured polychromatic linear attenuation values. The BH artifacts correction term is then derived by simulating the polychromatic and the monochromatic projections using an in-house implemented GPU-based MC model. The polychromatic projection is estimated accurately utilizing the LSE method, while the monochromatic projection is estimated by selecting the energy bin which produces the lowest MSE with the acquired projection. Unlike most of the available BH correction methods, the proposed one is fast and does not require the prior knowledge of the materials. Moreover, as the accuracy of the BH correction highly depends on the estimation accuracy of the polychromatic projection, it is shown that the proposed estimation method can predict a precise polychromatic projection that is very close to the one from the scanner. Thus, a successful non-linearity adjustment of the original corrupted projections has been achieved which allows the significant suppression of BH artifacts. We have validated the proposed BH correction method with the representative empirical BH correction method. Experimental results show that the proposed method outperforms the empirical method as the streak and cupping artifacts were significantly suppressed. It is also shown that the proposed method is robust against segmentation errors and scattering correction is introduced to improve BH correction results. This work shows that the proposed method is very effective and robust in the removal of the BH artifacts without previous knowledge of the materials which makes it very promising for medical and industrial applications.

\section*{Acknowledgment}

This work was supported by the German Academic Exchange Service (DAAD, No. 57381412) and the German Research Foundation (DFG, Germany) under the DFG-project
SI 587/18-1 in the priority program SPP 2187.



%





\ifCLASSOPTIONcaptionsoff
  \newpage
\fi

\bibliographystyle{IEEEtran}
{\footnotesize
\bibliography{IEEEabrv,bare_jrnl}}

\begin{thebibliography}{10}
\providecommand{\url}[1]{#1}
\csname url@samestyle\endcsname
\providecommand{\newblock}{\relax}
\providecommand{\bibinfo}[2]{#2}
\providecommand{\BIBentrySTDinterwordspacing}{\spaceskip=0pt\relax}
\providecommand{\BIBentryALTinterwordstretchfactor}{4}
\providecommand{\BIBentryALTinterwordspacing}{\spaceskip=\fontdimen2\font plus
\BIBentryALTinterwordstretchfactor\fontdimen3\font minus
  \fontdimen4\font\relax}
\providecommand{\BIBforeignlanguage}[2]{{%
\expandafter\ifx\csname l@#1\endcsname\relax
\typeout{** WARNING: IEEEtran.bst: No hyphenation pattern has been}%
\typeout{** loaded for the language `#1'. Using the pattern for}%
\typeout{** the default language instead.}%
\else
\language=\csname l@#1\endcsname
\fi
#2}}
\providecommand{\BIBdecl}{\relax}
\BIBdecl

\bibitem{Boas}
F.~Boas and D.~Fleischmann, ``Computed tomography artifacts: {C}auses and
  reduction techniques,'' \emph{Imaging in Medicine}, vol.~4, no.~2, pp.
  229--240, 2012.

\bibitem{Gompel}
G.~Gompel, K.~Slambrouck, M.~Defrise, K.~Batenburg, J.~deMey, J.~Sijbers, and
  J.~Nuyts, ``Iterative correction of beam hardening artifacts in {CT},''
  \emph{Phys. Med. Biol.}, vol.~38, no.~7, 2011.

\bibitem{Krumm}
M.~Krumm, S.~Kasperl, and M.~Franz, ``Reducing non-linear artifacts of
  multi-material objects in industrial 3{D} computed tomography,'' \emph{NDT
  and E International}, vol.~41, no.~4, pp. 242--251, 2008.

\bibitem{Brooks}
R.~Brooks and G.~Chiro, ``Beam hardening in {X}-ray reconstructive
  tomography,'' \emph{Phys. Med. Biol.}, vol.~21, no.~3, pp. 390--398, 1976.

\bibitem{McDavid}
W.~McDavid, R.~Waggener, W.~Payne, and M.~Denis, ``Correction for spectral
  artifacts in cross-sectional reconstruction from {X}-rays,'' \emph{Med.
  Phys.}, vol.~4, no.~1, pp. 54--57, 1997.

\bibitem{Casteele}
E.~Casteele, D.~Dyck, J.~Sijbers, and E.~Raman, ``A model-based correction
  method for beam hardening artefacts in {X}-ray microtomography,''
  \emph{Journal of X-ray Science and Technology}, vol.~12, pp. 43--57, 2003.

\bibitem{Gao}
H.~Gao, L.~Zhang, Z.~Chen, Y.~Xing, and L.~Shuanglei, ``Beam hardening
  correction for middle-energy industrial computerized tomography,'' \emph{IEEE
  Trans. Nucl. Sci.}, vol.~53, pp. 2796--2807, 2006.

\bibitem{Mou}
X.~Mou, S.~Tang, and H.~Yu, ``Comparison on beam hardening correction of {CT}
  based on {H-L} consistency and normal water phantom experiment,'' \emph{Proc.
  SPIE}, 2006.

\bibitem{Reiter}
M.~Reiter, F.~Oliveira, M.~Bartscher, C.~Gusenbauer, and J.~Kastner, ``Case
  study of empirical beam hardening correction methods for dimensional {X}-ray
  computed tomography using a dedicated multi-material reference standard,''
  \emph{Journal of Nondestructive Evaluation}, vol.~38, no.~10, 2018.

\bibitem{Alsaffar}
A.~Alsaffar, S.~Kie{\ss}, K.~Sun, and S.~Simon, ``Computational scatter
  correction for high-resolution flat-panel {CT} based on a fast {M}onte
  {C}arlo photon transport model,'' \emph{pre-print arXiv}, 2022,
  2201.13191,eess.IV.

\bibitem{Millner}
M.~Millner, W.~Payne, R.~Waggener, W.~McDavid, M.~Dennis, and V.~Sank,
  ``Determination of effective energies in {CT} calibration,'' \emph{Med.
  Phys.}, vol.~5, no.~6, 1978.

\bibitem{Hsieh}
J.~Hsieh, ``Computed tomography: principles, design, artifacts, and recent
  advances,'' 2012.

\bibitem{Hunter}
A.~Hunter and W.~McDavid, ``Characterization and correction of cupping effect
  artefacts in cone beam {CT},'' \emph{Dentomaxillofacial Radiol}, vol.~41,
  no.~3, pp. 217--223, 2012.

\bibitem{Thomsen}
M.~Thomsen \emph{et~al.}, ``Prediction of beam hardening artefacts in computed
  tomography using {M}onte {C}arlo simulations,'' \emph{Nuclear Instruments and
  Methods in Physics Research Section B: Beam Interactions with Materials and
  Atoms}, vol. 342, no.~1, pp. 314--320, 2015.

\bibitem{Alvarez}
R.~Alvarez and A.~Macowski, ``Energy-selective reconstructions in x-ray
  computerized tomography,'' \emph{Phys. Med. Biol}, vol.~21, pp. 733--744,
  1976.

\bibitem{Macovski}
R.~Macovski \emph{et~al.}, ``Energy dependent reconstruction in {X}-ray
  computerized tomography,'' \emph{Comp. Bol. Med.}, vol.~6, pp. 325--336,
  1976.

\bibitem{Hammersberg}
P.~Hammersberg and M.~Mangard, ``Correction for beam hardening artefacts in
  computerised tomography,'' \emph{Journal of X-ray Science and Technology},
  vol.~8, no.~1, pp. 75--93, 1998.

\bibitem{Herman}
G.~Herman, ``Correction for beam hardening in computed tomography,''
  \emph{Physics in Medicine and Biology}, vol.~24, pp. 81--106, 1979.

\bibitem{Nalcioglu}
O.~Nalcioglu and R.~Lou, ``Post-reconstruction method for beam hardening in
  computerized tomography,'' \emph{Phys. Med. Biol.}, vol.~24, no.~2, 1979.

\bibitem{Zhao}
W.~Zhao, D.~Li, K.~Niu, W.~Qin, H.~Peng, and T.~Niu, ``Robust beam hardening
  artifacts reduction for computed tomography using spectrum modeling,''
  \emph{IEEE Transactions on Computational Imaging}, vol.~5, no.~2, pp.
  333--342, 2019.

\bibitem{Elbakri}
I.~Elbakri and J.~Fessler, ``Statistical image reconstruction for
  polyen-ergetic x-ray computed tomography,'' \emph{IEEE Trans. Med. Imag.},
  vol.~21, no.~2, pp. 89--99, 2002.

\bibitem{Elbakri2}
I.~Elbakri \emph{et~al.}, ``Segmentation-free statistical image reconstruction
  for polyenergetic {X}-ray computed tomography with experimental validation,''
  \emph{Phys. Med. Biol.}, vol.~48, no.~15, pp. 2453--2477, 2003.

\bibitem{Cai}
C.~Cai \emph{et~al.}, ``A full-spectral bayesian reconstruction approach based
  on the material decomposition model applied in dual-energy computed
  tomography,'' \emph{Med. Phys.}, vol.~40, no.~11, pp. 1--16, 2013.

\bibitem{Wu}
M.~Wu \emph{et~al.}, ``A practical statistical polychromatic image
  reconstruction for computed tomography using spectrum binning,'' \emph{Proc.
  SPIE Med. Imag.}, vol. 9033, no.~90, 2014.

\bibitem{Yang}
Q.~Yang \emph{et~al.}, ``Evaluation of spectrum mismatching using spectrum
  binning for statistical polychromatic reconstruction in {CT},''
  \emph{Springer}, pp. 42--47, 2014.

\bibitem{Gu}
R.~Gu and A.~Dogandzic, ``Blind beam-hardening correction from poisson
  measurements,'' \emph{Nuclear Instruments and Methods in Physics Research A
  349}, 2016.

\bibitem{Yan}
C.~Yan, R.~Whalen, G.~Beaupre, S.~Yen, and S.~Napel, ``Reconstruction algorithm
  for polychromatic {CT} imaging: Application to beam hardening correction,''
  \emph{IEEE Trans. Med. Imag.}, vol.~19, no.~1, pp. 1--11, 2000.

\bibitem{Man}
B.~De~Man \emph{et~al.}, ``An iterative maximum-likelihood polychromatic
  algorithm for {CT},'' \emph{IEEE Trans. Med. Imag.}, vol.~20, no.~10, pp.
  999--1008, 2001.

\bibitem{Menvielle}
N.~Menvielle, Y.~Goussard, D.~Orban, and G.~Soulez, ``Reduction of beam
  hardening artifacts in x-ray {CT},'' \emph{IEEE Eng. Med. Biol. 27th Annu.
  Conf.}, pp. 1865--1868, 2005.

\bibitem{Brabant}
L.~Brabant \emph{et~al.}, ``A novel beam hardening correction method requiring
  no prior knowledge, incorporated in an iterative reconstruction algorithm,''
  \emph{NDT E Int.}, vol.~51, pp. 68--73, 2012.

\bibitem{Jin}
P.~Jin, C.~Bouman, and K.~Sauer, ``A model-based image reconstruction algorithm
  with simultaneous beam hardening correction for x-ray {CT},'' \emph{IEEE
  Trans. Comput. Imag.}, vol.~1, no.~3, pp. 200--216, 2015.

\bibitem{Schumacher}
D.~Schumacher, R.~Sharma, J.~Grager, and M.~Schrapp, ``Scatter and beam
  hardening reduction in industrial computed tomography using photon counting
  detectors,'' \emph{Meas. Sci. Technology}, vol.~29, 2018.

\bibitem{GEANT4}
S.~Agostinelli \emph{et~al.}, ``{GEANT}4: {A} simulation toolkit,'' \emph{Nucl.
  Instrum. Meth.}, pp. 250--303, 2003.

\bibitem{Otsu}
N.~Otsu, ``A threshold selection method from gray-level histograms,''
  \emph{IEEE Transactions on Systems, Man, and Cybernetics}, vol.~9, no.~1, pp.
  62--66, 1979.

\bibitem{SpekCalc}
G.~Poludniowski \emph{et~al.}, ``Spek{C}alc: a program to calculate photon
  spectra from tungsten anode x-ray tubes,'' \emph{Phys. Med. Biol.}, vol.~54,
  no.~19, 2009.

\bibitem{Kyriakou}
Y.~Kyriakou, E.~Meyer, D.~Prell, and M.~Kachelriess, ``Empirical beam hardening
  correction (ebhc) for {CT},'' \emph{Med. Phys.}, vol.~37, pp. 5179--5187,
  2010.

\end{thebibliography}

\end{document}